\documentclass[twocolumn,aps,prb]{revtex4-1}
\usepackage{amsmath}
\usepackage{epsfig,color}
\usepackage{graphicx}
\usepackage{dcolumn}
\usepackage{bm}
\usepackage{multirow}
\usepackage{bm}
\usepackage{enumerate}
\usepackage{graphicx}
\usepackage{verbatim}
\usepackage{hyperref}
\hypersetup{colorlinks=true,linkcolor=blue,anchorcolor=blue,citecolor=blue,filecolor=blue,urlcolor=blue,bookmarksnumbered=true,pdfview=FitB}
\usepackage[utf8]{inputenc}
\usepackage[english]{babel}

\newcommand{\ve}{\varepsilon}

\newcommand{\bk}{{\bf k}}
\newcommand{\bp}{{\bf p}}
\newcommand{\bq}{{\bf q}}

\newcommand{\nn}{\nonumber}
\newcommand{\beq}{\begin{equation}}
\newcommand{\eeq}{\end{equation}}
\newcommand{\bea}{\begin{eqnarray}}
\newcommand{\eea}{\end{eqnarray}}
\newcommand{\bse}{\begin{subequations}}
\newcommand{\ese}{\end{subequations}}
\newcommand{\bwt}{\begin{widetext}}
\newcommand{\ewt}{\end{widetext}}

\newcommand{\bkp}{{\bf k}'}

\newcommand{\bv}{{\bf v}}

\newcommand{\bsu}{\begin{subequations}}
\newcommand{\esu}{\end{subequations}}

\newcommand{\bQ}{{\bf Q}}

\begin{document}

\title{
Gradient terms in quantum-critical theories of itinerant fermions}
\author{
Dmitrii L. Maslov$^{a,b}$, Prachi Sharma$^{a}$, Dmitrii Torbunov$^c$, and Andrey V. Chubukov$^{c}$}
\date{\today}
\affiliation{
$^{a)}$Department of Physics, University of Florida, Gainesville, FL 32611-8440, USA\\
$^{b)}$National High Magnetic Field Laboratory, Tallahassee, Florida 32310, USA\\
$^{c)}$Department of Physics, University of Minnesota, Minneapolis, MN 55455, USA}

\date{\today}

\begin{abstract}
We investigate the origin and
renormalization
 of the gradient ($Q^2$) term in the propagator of soft bosonic fluctuations in theories of itinerant fermions near  a quantum critical point (QCP) with $Q =0$.
A common
  belief is that (i) the $Q^2$  term comes from fermions with high energies (roughly of order of the bandwidth)
  and,
   as such,
   should be included into the bare bosonic propagator of the effective low-energy  model,
   and (ii) fluctuations within the low-energy model generate Landau damping of  soft bosons, but
 affect the $Q^2$ term
only weakly.
 We argue that the situation is
in fact
more
complex.
   First, we found that
  the
  high-
  and low-energy contributions to
  the
  $Q^2$ term
  are of the same order.
 Second, we
  computed
  the
  high-energy contributions to
  the
  $Q^2$
  term
  in two microscopic models
  (a Fermi gas with Coulomb interaction and
    the
  Hubbard model)
  and found that in all cases these contributions are numerically much smaller than the
  low-energy ones, blue especially in 2D.
  This last result is relevant for the behavior of observables at low energies,
   because
   the low-energy part of the  $Q^2$ term
    is expected to flow when the effective mass diverges near QCP.
    If this term is the dominant one, its flow has to be computed self-consistently,
    which
    gives rise to a novel
    quantum-critical
    behavior.   Following up on these results,
we discuss two possible ways of formulating the
theory
of a
 QCP with $Q=0$.
  \end{abstract}

\maketitle

\section{Introduction}
\label{sec:intro}
Understanding the behavior  of itinerant fermions  near a quantum critical point (QCP) is crucial for describing correlated electron systems. The
underlying idea is that near QCP the behavior becomes universal and can be described by a small number of exponents, which depend only
on the type of symmetries broken in the ordered phase,
   and on spatial dimensionality.

\begin{figure}[htb]
    \centering
    \includegraphics[width=1.0\columnwidth]{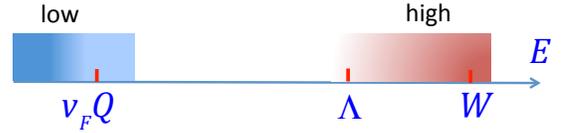}
    \vspace{-1.2in}
    \caption{Energy scales contributing to the gradient term of a quantum-critical theory. $Q$ is the external momentum (for a finite-$Q$ QCP, it is to be understood as $|\bQ-\bQ_0|$, where $\bQ_0$ is the ordering momentum), $\Lambda$ is the cutoff of the low-energy theory, and $W$ is the bandwidth.
    The region $E\lesssim v_FQ$ is referred to as ``low energies" and the region $\Lambda\lesssim E \lesssim W$ as to ``high energies''.}
    \label{fig:scales}
\end{figure}

  QCP
 in a metallic system
   generally occurs
   at
    intermediate coupling, where
    the
     fermion-fermion interaction is
      of
      order of
      bandwidth $W$.  In this case,
      a
      perturbation theory in the
      original
      interaction is not
       a reliable computational scheme.
        A commonly accepted alternative
        \cite{hertz:1976,millis:1993,moriya:1985,abanov:2003}
     is to
      abandon the underlying microscopic model and
       analyze instead an effective low-energy model of fermions interacting
       via
       the
       exchange
       of
       soft
       bosons
       that
       condense in the ordered state.
       Although such a model cannot,  except for a few special cases, be derived in a controllable way from
      microscopics,
      it is generally  believed to emerge once one integrates out fermions with energies between $W$ and some much smaller energy $\Lambda$, which serves as the upper cutoff for the effective model
       (see Fig.~\ref{fig:scales}).
      The anticipated universality of low-energy behavior implies that the behavior of observables at large distances and
      long
      time scales
       does not depend on
      a particular choice of $\Lambda$,
      as long as $\Lambda/W \ll 1$.

        The inputs for the low-energy model are
        the
        boson-fermion coupling and
        bare bosonic propagator, $\chi_0 (\bQ, \Omega)$.   The latter is a particle-hole polarization bubble,  dressed by interactions involving fermions with energies between $\Lambda$ and $W$. Fermions with such energies  are  assumed  not to differ
         qualitatively from free  one, even at QCP. As the consequence, the momentum and  frequency dependences of $\chi_0 (\bQ, \Omega)$  are assumed to be regular, i.e., expandable
          in powers of $\Omega^2$ and $(\bQ-\bQ_0)^2$, where $\bQ_0$ is the momentum at which bosons condense
          in the ordered phase.
           The regular frequency dependence of $\chi_0 (\bQ, \Omega)$ is often
           omitted in anticipation that renormalizations within the low-energy model produce a much stronger,
            non-analytic
             $|\Omega|\Gamma(\bQ)$ frequency dependence
           (Landau damping). This  non-analytic dependence comes from fermions with energies
           $\lesssim |\Omega|$,  and  it emerges
           regardless of
            whether
            the system is at QCP or away from it.
              With few exceptions (discussed later in this section),\cite{belitz:2005,woelfle:2007,brando:2016}
               there is no analogous non-analytic contribution
               to
               the  momentum dependence of the bosonic propagator,  hence the dependence of $\chi_0$ on $(\bQ - \bQ_0)^2$ is essential and must be kept.  In most  theories, this dependence is  taken to
            be of the
             Ornstein-Zernike form:
         \beq
        \chi_0 (\bQ, \Omega) \approx \chi_0 (\bQ, 0) = \frac{\chi_0}{c^2(\bQ-\bQ_0)^2+M^2},
        \label{a_1}
        \eeq
        where $\chi_0 $
        is
         of order of the static
        and
        uniform susceptibility of free fermions,
         $c$ is a model-dependent parameter, generally of order of the interatomic spacing, and
        (dimensionless) $M$ is
        the
        measure of
        the
        distance to QCP.
        Boson-fermion
        models with $\chi_0 (\bQ, \Omega)$ given by Eq.~(\ref{a_1}) have been
        studied extensively
        both for
        finite $\bQ_0$ (a density-wave QCP) and
         $\bQ_0=0$ (a ferromagnetic or nematic QCP, or the model of fermions interacting with a gauge field).   When only
         the
         Landau-damping contribution from low-energy fermions is included,
         the
         critical bosonic propagator has
         dynamical exponent
         of
         $z=2$
          for
         a finite-$\bQ_0$ QCP
         [in this case, the prefactor $\Gamma (\bQ)$ of the Landau-damping term is
         a
         constant]
         and $z=3$
         for
          a $\bQ_0=0$ QCP
          [in this case, $\Gamma(\bQ)=f(\hat \bQ)\gamma/Q$ with $\hat \bQ\equiv \bQ/Q$].  Beyond this approximation,
          the
          interaction between low-energy fermions and critical bosonic fluctuations gives rise to
          a
          singular frequency dependence of the fermionic self-energy in dimensions $d\leq 3$,
          leading to a non-Fermi liquid (NFL)  behavior at QCP. In 2D (and in some specially crafted
          models
           in $2< d <3$) these singular renormalizations also give rise to anomalous exponents  for fermionic and bosonic propagators, and may also change the value of the
            dynamical exponent $z$.

        In this paper,
        we discuss another aspect of the low-energy model,
        which attracted less attention until
        recently.\cite{wolfle:2011,abrahams:2014}
   This new aspect
         is an observation that
        low-energy fermions
        can
        also contribute to
        the
        regular $(\bQ-\bQ_0)^2$ dependence of the bosonic propagator.  This  additional contribution is often neglected because it is assumed  to be smaller than
        that
        from high-energy fermions by
        a factor of
        $\Lambda/W\ll 1$,  simply because the energy width
        of
        the low-energy model is small
        (below we
        will
        dispute this assumption for
        the
        ${\bQ_0} =0$ case).  However, the parameter $c$ in
        the high-energy contribution
        [Eq.~(\ref{a_1})] is model-dependent
        and, in principle, can be rather small.  If we assume momentarily that this is the case and set $c=0$ in Eq.~(\ref{a_1}), we
        encounter
        a situation when the bare bosonic propagator is just a constant, and both the Landau-damping and
        gradient
        terms
         in
         $\chi (\bQ, \Omega)$ come from low-energy fermions.  Away from
         a
         QCP, it does not really matter
         whether the
         gradient terms come from high or low energies.
         Near
         a
         QCP, however, the low-
         and high-energy contributions are
         qualitatively different.  Namely, the
         high-energy
         contribution
          is insensitive to NFL physics,
          specifically
          to the divergence of the effective mass.
          The contribution from low-energy fermions, on the other hand, depends on
          the effective mass.
          In Eliashberg-type theories, where the self-energy near
          a
          QCP
          depends
          predominantly
          on the frequency,
                   the fermionic
          residue $Z (\omega) = \left[1
           -
           \partial\Sigma (\omega)/\partial \omega)\right]^{-1}$ accounts for
           renormalization of the fermionic dispersion [$\ve_\bk \to Z (\omega)\ve_\bk]$,
           which means that the ratio of the effective and bare quasiparticle masses is $m^*/m = 1/Z$.  Then, if one keeps only the low-energy contribution to
           the gradient
             term, one ends up with a new theory in which mass renormalization has to be computed self-consistently
              with the bosonic dispersion.
              We emphasize that this holds even in cases
             when
             bosonic propagators do not acquire
             anomalous dimensions.

        The
        spin-fermion model with
        an
         additional
          mass-dependent prefactor
          of the gradient term
          in the bosonic
          propagator has been recently put forward by Wolfle and Abrahams\cite{wolfle:2011}
          in the context of an antiferromagnetic QCP in $d=3$.
           In this case a high-energy contribution to
           the gradient term
           should
           also
           be present, and it is
           not
           {\em a priori}
            clear why it can be neglected, given that in a generic case this  should be the largest contribution.  At the same time, the analysis in Ref.~\onlinecite{wolfle:2011} and
            subsequent work
            \cite{abrahams:2014}
            demonstrated a very good agreement between
            this
            theory and the data for
            a number of
            heavy-fermion materials. This calls for further investigation of the interplay between high-
            and low-energy contributions to the gradient term
            in these
            systems.

        In this paper, we study the case
        of
        $\bQ_0=0$,
        which
         is special for two reasons. First, we show
          that low-energy contribution to the ${Q}^2$  is of the same order as high-energy contribution, i.e., it is not small in $\Lambda/W$.  Second, we show that the high-energy
           contribution to the ${Q}^2$ term is absent if
          the interaction
           between  high-energy fermions
           is approximated as static,
         and
         emerges
         only
          if
          one includes
          dynamical screening of
           this  interaction.

To demonstrate these two features, we analyze the interplay between  high-
and low-energy gradient terms first within
the
random phase approximation (RPA), which neglects dynamical screening of the interaction by high-energy fermions,  and then for two  models with a dynamical interaction between high-energy fermions.

 The first  model is a Fermi gas with a dynamically screened Coulomb interaction.
In principle,
this
model can be tuned to a critical point in the spin channel
or in the charge channel
 with angular momentum $l\geq 2$.   We will not analyze
a
 specific path to quantum criticality, but rather compute the $Q^2$ term in
 dressed bosonic
 propagators in the spin and charge channels.
   We show that the low-energy contribution
 to the $Q^2$ term comes from fermions with energies  of order of $v_F Q$, where $v_F$ is the Fermi velocity,
 while
 the
   high-energy contribution
   comes from energies of order of the
 effective
 plasma frequency, $\Omega_p$.
 Our reasoning for the separation between
 the
  low-
 and high-energy contributions holds if $\Omega_p
 \geq \Lambda$, which in practical terms implies that the enhancement of the mass ratio $m^*/m$  is confined
 to energies below
 $\Lambda$.
 We show that the
 (numerically)
  dominant contribution to
 the ${Q}^2$ term
  comes from low-energy fermions
 both in 2D and 3D.
 The difference between the high-
 and low-energy contributions is particularly spectacular in 2D, where  the high-energy contribution accounts only for  two
  percent of the total.
         Moreover, we found that the sign of the high-energy part of the $Q^2$ term is non-universal:
          it is negative in 2D and  positive  in 3D for fermions with a parabolic dispersion. For 2D fermions on a square lattice
          the high-energy part of the $Q^2$ term changes sign
          near quarter-filling and stays positive all the way up to half-filling.

  The second model is a Fermi gas with a parabolic dispersion and
  Hubbard interaction,
   which
      we set to be a constant
      ($=U$)
    up to some  momentum cutoff
    and then vanish.
  This model can be tuned to QCP in the spin channel (for positive $U$) or the charge channel (for negative $U$). We show that the high-energy contribution to
  the $Q^2$ term emerges at second order in $U$,
  once dynamical screening of the Hubbard interaction is included.
  As in the Coulomb case,
  we find that high-energy contribution
  to the $Q^2$ term is numerically small.
  We extended the result beyond second order in $U$  by summing up
  RPA diagrams for screened interaction,
  and found that
   the prefactor of the high-energy $Q^2$ term changes sign at some critical value of $U$.

        The outcome of our analysis is that the low-energy boson-fermion model for QCP with $\bQ_0 =0$ has to be reconsidered, at least in some cases. Namely, instead of
        starting from Eq.~(\ref{a_1}) for the bare propagator and neglecting additional $Q^2$ contributions to $\chi (\bQ, \Omega)$ from low-energy fermions, one has to set the bare propagator to be $\bQ$-independent and compute both the frequency
        and
        momentum dependences of $\chi (\bQ, \Omega)$ within the low-energy model. The Landau-damping term does not depend on $m^*/m$ and is the same as for free fermions, but the prefactor of the $Q^2$ term
        is reduced by critical fluctuations.  As
        a consequence,
        the
        quantum-critical theory becomes qualitatively different from the one
        with
        the bare propagator
        given by Eq.~(\ref{a_1}).

        We caution that,
        near
        QCP, our arguments
        apply
        to the behavior of the system
        at small but finite energies. At progressively smaller energies, the low-energy contribution to the $Q^2$ term  is
        reduced by growing $m^*/m$  and eventually gets
        smaller than the high-energy contribution, which is
        not affected by mass renormalization.
        As a result,
         the high-energy contribution dominates at the lowest energies, and the
         quantum-critical
           theory
           eventually
           becomes the "conventional" one, if the high-energy $Q^2$ term is positive.  If it is negative,
           the system either develops an incommensurate order or the transition becomes first order.
           We emphasize that this is different from a first-order transition and
        an
        incommensurate magnetic order due to generation of a non-analytic momentum dependence
         of
          $\chi (\bQ, 0)$
         by
         an effective
         long-range
         interaction.\cite{belitz:2005,woelfle:2007,brando:2016}
           The effect we consider here is related to
         a
         possible sign change of the
           analytic
            $Q^2$ term.
          One
        difference is that
         the  effect due to non-analyticity
        holds for an
        $O(3)$-symmetric ferromagnetic QCP, but is absent for a charge QCP and also  if
        the $O(3)$ symmetry is broken down to
        Ising\cite{chubukov:2004,rech:2006} by, e.g., spin-orbit interaction.\cite{zak:2010,zak:2012}
        In contrast, the new physics, associated with potential negative sign of the $Q^2$ term, holds for both spin and charge QCPs.
        Another difference is that a non-analytic $Q$-dependence of   $\chi (\bQ, 0)$
 is a low-energy effect, while we are interested in a high-energy $Q^2$ term.

The rest of the paper is organized as follows.
 In Sec.~\ref{RPA} we analyze the  dressed bosonic propagator near $\bQ=0$ within RPA and FL theory.
In  Sec.~\ref{sec:RPA2} we show that
 the $Q^2$ term in RPA comes exclusively from fermions with energies  of order of $v_F Q$,
 while the high-energy contribution
  is absent. In Sec.~\ref{sec:FL} we
  include FL renormalizations on top of RPA.   In Sec.~\ref{sec:coulomb} we compute both the low-
  and high-energy contributions to
  the
  $Q^2$ term in the bosonic propagator for a model with dynamically screened  Coulomb interaction.
  We show that the low-energy contribution still comes
  from
  energies  of order of $v_F Q$, while the high-energy
  one comes from energies of order of the effective plasma frequency.  In Sec.~\ref{sec:Hubbard} we
  perform the same analysis for the Hubbard model.
    We discuss
  possible
  consequences of our results for low-energy
  theories
  of a ${\bf Q}_0=0$ QCP in Sec. \ref{sec:analysis}.
  Technical details of the calculations are given in Appendices \ref{sec:dia}-\ref{sec:static}.

\section{
Bosonic propagator in RPA and
in
FL theory}
\label{RPA}
\subsection{RPA}
\label{sec:RPA2}

To illustrate the issue with the gradient term in the bosonic propagator near the $\bQ_0 =0$ criticality, we first  consider
derivation of Eq.~(\ref{a_1})
 within RPA
   for a system with a constant repulsive interaction $U$.
 A system with sufficiently large repulsive $U$ is unstable towards ferromagnetism, and we focus on the spin susceptibility.

 For free fermions, the static spin susceptibility $\chi^s(\bQ)=\Pi (\bQ)$, where
 $\Pi (\bQ)$ is the free static polarization bubble  (with an extra factor of two due to spin summation)
\beq
\Pi(\bQ) =-2 \int \frac{d^dk}{(2\pi)^d} \int^\infty_{-\infty} \frac{d\omega_m}{2\pi}G(\bk+\bQ,\omega_m) G(\bk,\omega_m),
\label{pifree}
\eeq
 where $G(\bk,\omega_m)=(i\omega_m-\ve_\bk)^{-1}$ is the Green's function and $\omega_m$ is the Matsubara frequency.
  The  dressed
  spin susceptibility is given by a series of ladder
  diagrams, which is summed up into
 \beq
 \chi^s(\bQ)= \frac{\Pi(\bQ)}{1 - \frac{U}{2} \Pi(\bQ)}.
 \label{chi_1}
 \eeq
In the limit $\bQ \to 0$  the polarization bubble is reduced to $\Pi (\bQ \to 0)=N_F$, where $N_F$ is the
 density of states at a Fermi level  per two spin orientations.
At finite
$Q$, $\Pi (\bQ)$ should in general have some regular dependence on $\bQ$, i.e., to be expandable in powers of $Q^2$:
\beq
\Pi(\bQ)= N_F  + A \frac{Q^2}{k^2_F} + {\cal O}(Q^4).
\label{n_1}
\eeq
The prefactor $A$ vanishes in special cases, e.g.,
for 2D fermions with a parabolic  or linear dispersion, but in general is non-zero.
 The issue we consider is
 where this term comes from.

The constant ($N_F$) term in Eq.~(\ref{n_1}) can be obtained in two ways -- by integrating first over frequency and then over the momentum
or vice versa.\cite{maslov:2010}
In the first method, one has to keep $Q$ finite and set it to zero only at the end of calculation. The integral comes from the region where the poles
 of the integrand over $\omega_m$ in  Eq.~(\ref{pifree}) are in the opposite half-planes. This imposes the conditions $\ve_{\bk+\bQ}>0$ and
$\ve_{\bk}<0$ (or {\em vice versa}) which, for $Q\ll k_F$, can be satisfied only for $\bk$ near the Fermi surface,  when $\ve_\bk$ is at most comparable to $v_F Q$. In our nomenclature, this implies that the integral comes from low energies.
In the second method, one can set $Q=0$ from the beginning but
constraints
 integration over  $\ve_\bk$
 to the region
  $-W \leq \ve_\bk \leq W$.   From Eq. (\ref{pifree}) we then have
\bea
\Pi(0) &=& -N_F \int^\infty_{-\infty}  \frac{d\omega_m}{2 \pi} \int_{-
W}^
W d \ve_\bk \frac{1}{(i \omega_m - \ve_\bk)^2} \nonumber \\
&& = N_F \int^\infty_{-\infty} \frac{d\omega_m}{2 \pi} \frac{2
W}{\omega_m^2 +
W
^2} =N_F.
\eea
This time, the integral comes from energies $|\omega_m| \sim |\ve_\bk| \sim
W$, i.e., from high energies in our nomenclature.  The fact that the same result can be obtained in two
ways implies that the $N_F$ term in the polarization operator for free fermions is an "anomaly",\cite{treiman:1972,chubukov:2008} which  can be  viewed as
either as
a
low- or high-energy contribution,
depending on the regularization procedure.  This feature is
a
consequence of
the double pole in the integrand
of  Eq.~(\ref{n_1}) for $\bQ=0$.

For the $Q^2$ term,  the situation is different.  If we compute this term by the second method, i.e., by expanding the integrand in $\Pi (\bQ)$  to order $Q^2$
and integrate first over $\ve_\bk$ in finite limits $-
W \leq \ve_\bk \leq
W
$ and then over frequency, we get zero.  This implies that there is no "high-energy" contribution to $\Pi (\bQ)$  for free fermions.  If, on the other hand, we keep $\bQ$ finite
and integrate over frequency first, we do find a non-zero $Q^2$ term. An explicit calculation for 3D fermions with an isotropic but otherwise arbitrary
 dispersion $\epsilon_{k}\equiv\ve_\bk+E_F$
 yields
 \cite{maslov:2009}
\bea
A &=&- \frac{k_{F}^{2}}{12}\frac{d}{d\epsilon_k}\left\{ N (\epsilon_k) \left[
\frac{2
v_k}{k}+\frac{1}{m_k}\right]\right.\nn\\
&&\left.-\frac{2}{3}\frac{d}{d\epsilon_k }\left[ N(\epsilon_k) v_k^{2}\right]
\right\}\Big|_{\epsilon_k =E_F},
\label{A}
\eea
where $v_k=d\epsilon_k /dk$,
$m_k^{-1}=d^{2}\epsilon_k /dk^{2}$, and $
N(\epsilon_k)$ is the
density of states as a function of
energy.
For a power-law dispersion, $\epsilon_k = ak^{\gamma }$, we have $
A =-(\gamma +1)N_F/36$.
For a parabolic dispersion
$A =-N_F/12$.   A similar formula holds for
the
2D case, the only difference is that
$A=0$ in 2D
both
for
parabolic and linear
dispersions. The low-energy nature of this $Q^2$ term is manifested by the fact that its prefactor is expressed entirely via the dispersion and its derivatives at the Fermi energy,
i.e.,
$A$ comes from fermions with energies smaller than $v_F Q$.  In this respect, if we would construct the bare spin susceptibility for the effective low-energy model  by
integrating out
 fermions with energies much larger than $v_F Q$, we would not obtain
 a $Q^2$ term.
 At the same time, we see that
   the prefactor $A$ does not depend on the upper cutoff $\Lambda$
   of
    the low-energy model and, hence, does not contain
    a
    small prefactor
    of
    $\Lambda/W$.
(Following along the same lines,
  we show  in Appendix \ref{sec:dia} that the diamagnetic susceptibility of a free electron gas, which is usually viewed as the property of the entire electron band, is in
fact also a low-energy property in the sense defined above.)

\subsection{Gradient term within the FL theory}
\label{sec:FL}

 The computational procedure in which the constant $N_F$  term in the polarization bubble comes from
 low-energy fermions
 can be extended in a rigorous   way to include FL renormalizations.
 One way to
 to do this is to solve the kinetic equation for a FL in the presence of a
 magnetic field;\cite{agd:1963,nozieres:1966,lifshitz:1980} another is to keep with diagrammatics,\cite{finkelstein:2010,chubukov:2014} but to go beyond RPA and include
 self-energy and vertex corrections.
 Both procedures lead to the familiar result for the static and uniform spin susceptibility of a FL
\beq
\chi^s
= \frac{N^*_F}{1 + F_0^{a}}.
\label{chiFL1}
\eeq
 Within  diagrammatics, this result comes about because
self-energy corrections change the  low-energy part of the Green's function
 to
 $Z/(i\omega - \ve^*_\bk)$, where  $\ve^*_\bk = v^*_F (k-k_F)$,
 $v^*_F = k_F/m^*$,
 and
 $m^*$ is the effective mass.
The role of vertex corrections is to cancel the $Z$ factors coming from the numerators of the Green's functions. Also, the constant interaction $U$ is replaced by
the zeroth harmonic of the Landau function in the spin channel FL  via $Z^2UN^*_F\to-F_0^a$, where $N_F^*=N_F(m^*/m)$ is the renormalized density of states at the Fermi level.

How FL renormalizations affect
the $Q^2$ term
 is a more
 difficult
  question,
  which,
 in general,
has no
 definite
  answer
in either the kinetic-equation or diagrammatic versions of the FL theory.
Indeed, the FL theory operates with quasiparticles with dispersions linearized near the Fermi energy and thus contains only the first derivative of the dispersion
(the Fermi velocity) but not higher derivatives, whereas one needs to know higher derivatives
 of the dispersion
   to obtain a $Q^2$ term in the susceptibility
   [see Eq. (\ref{A})].
 Keeping higher than ${\cal O}(k-k_F)$ terms in the dispersion is, strictly speaking, inconsistent with
  a FL
  assumption of non-decaying quasiparticles, because damping of
 quasiparticles occurs already at order $(k-k_F)^2$.

One can approximately
relate the prefactor of the $Q^2$ term
 to the renormalized effective mass (which is a FL parameter)
  in the case of a local FL, when
 the self-energy  depends on the frequency stronger than
 on the momentum.  Such a case is realized near
   a QCP with $z>1$
  (Refs.~\onlinecite{chubukov:2005self,metlitski:2010b,metlitski:2010c}).
     In this situation, fermionic propagator can be approximated by $G(\bk, \omega_m) = Z/(i \omega_m - Z \ve_\bk)$, i.e., the whole fermionic dispersion acquires a factor $Z$. One can  then re-calculate the $Q^2$ term  for free fermions with dispersion $Z \ve_\bk$ with an obvious result that $A$ in  Eq.~(\ref{A}) is multiplied by
$Z  = m/m^*$. (The overall factor of $Z$ in $G(\bk,\omega_m)$ is canceled by vertex corrections.)  Near
a
 QCP, $m^*/m$ is supposed to diverge and thus the $Q^2$ term
vanishes. Therefore, mass renormalization
 changes the critical theory in a qualitative way compared to the case when
  the prefactor
  of
  the $Q^2$ term
  is treated as
 a
 constant.

  Although we said that the $Q^2$ term, in general, cannot be obtained within the FL theory, it is still instructive to follow the consequences of including this term phenomenologically, as it is done, e.g.,  in some models of nematic instabilities.\cite{oganesyan:2001,wu:2004,wu:2007}
Consider a special case of a FL, in which  the interaction in the spin channel contains only the zeroth harmonic of the Landau function, $F_0^a$. Solving a FL kinetic equation \cite{agd:1963,nozieres:1966,lifshitz:1980}
in the presence of a time- and position-dependent magnetic field, we obtain for the spin susceptibility
\beq
\chi^s(\bQ,\Omega)=\frac{
  \Pi^*(\bQ,\Omega)}{1+F^a_0
  \Pi^*(\bQ,\Omega)/N_F^*},\label{chiFL1}
\eeq
where
\beq
\Pi^*(\bQ,\Omega)=N_F^*\left\langle\frac{\bv^*_F\cdot\bQ}{\bv^*_F\cdot\bQ-\Omega-i0^+}\right\rangle
\eeq
is  the particle-hole polarization bubble in the small-$Q$ limit,
  dressed by FL corrections,
   and $\langle\dots\rangle$ denotes averaging over the angle between $\bv_F^*$ and $\bQ$. In the limit of $\Omega\ll v_F^*Q$, Eq.~(\ref{chiFL1}) is reduced to\footnote{It can be shown \cite{zyuzin} that Eq.~(\ref{chiFL2}) is valid for an arbitrary Landau function with any number of harmonics but only in the limit $\Omega/v_F^*q\ll 1$.}
\beq
\chi^s(\bQ,\Omega)= \frac{N_F^*}{1+F_0^a+iF_0^a\alpha_d\frac{\Omega}{v_F^*Q}}, \label{chiFL2}
\eeq
where $\alpha_d$ is a numerical coefficient which depends on spatial dimensionality $d$.
Equation (\ref{chiFL2}) does contain
the bosonic mass (proportional to $1+F_0^a$) and the Landau-damping term but, as  expected, it does not have a $Q^2$ term. In this version of the FL theory, the $Q^2$ term
 is absent because the quasi-classical kinetic equation contains only the first gradient of the distribution function, and thus $\bQ$ enters only as
  $\bv^*_F\cdot\bQ$.\footnote{This is also the reason why the FL theory does not have analog of Eq.~(\ref{chiFL1})  for the {\em diamagnetic} susceptibility, which is
   obtained as the prefactor of the $Q^2$ term in the static current-current correlation function (see Appendix \ref{sec:dia}). For
an explicit calculation of the diamagnetic susceptibility renormalized by the Coulomb interaction, see  Ref.~\onlinecite{vignale:1988} and references therein.}

Suppose now that we ``improve'' the theory
by
 adding
 a
 low-energy term $Q^2$ term to the polarization bubble:
 \beq
\Pi^*(\bQ,\Omega)/N_F^*=1+i\alpha_d\frac{\Omega}{v_F^*Q} +B\frac{Q^2}{k_F^2}.\label{q2}
\eeq
(To ensure the ordering occurs at $Q=0$, we must also assume that  $B<0$.)
Near criticality, when $F_0^a\approx -1$, the spin susceptibility then becomes
\beq
\chi^s(\bQ,\Omega)=\frac{N_F}{M^2+\frac{m}{m^*}B(Q/k_F)^2-i\alpha_D\frac{\Omega}{v_FQ}}
\eeq
with $M^2\equiv (1+F_0^a)(m/m^*)$.
We see again if the  fermionic  mass diverges near criticality while $B$ stays constant, the gradient term vanishes. This
implies that the gradient term becomes a part of the low-energy theory and needs to be renormalized accordingly.

In the next two sections we show,  on examples of two microscopic models,
that there is a finite  contribution to the $Q^2$ term from high-energy fermions.
 Such
 contribution is not  affected by
 by the divergence of the effective mass
 and hence
 survives at QCP. We found, however, that
 the high-energy contributions
  are numerically very small,
  at least
  in the models we considered.
  Therefore, at least over some range of energies, the dominant contribution to the
  $Q^2$ term
   comes from low-energy fermions,
    and
  its prefactor does
  contain $m/m^*$.

\section{Electron gas with Coulomb interaction}
\label{sec:coulomb}
\subsection{Formulation of the problem and background}
We consider 2D and 3D electron gases with Coulomb interaction in the high-density limit.
  As we have already said in Sec.~\ref{sec:intro}, a $Q^2$ term
  in
  the free-electron polarization bubble,
  $\Pi(\bQ,0)$, if non-zero,
 comes only from low
 energies.
 A $Q^2$ term in the bubble,
 renormalized by the dynamically screened Coulomb interaction,
   was calculated in a seminal 1968 paper
by Ma and Brueckner for a 3D electron gas.  \cite{ma:1968}  Since then, the $Q$-dependence
  has been
  addressed by a large number of authors by  semi-analytic or numerical means, see, e.g., Refs.~\onlinecite{geldart:1970,rajagopal:1977,maldague:1978,khalil:2002}.
The difference between our calculation and the previous ones is that the goal of the latter  was to obtain the entire $Q^2$ term, which contains both low- and high-energy parts.
 On the other hand,
 we
  are interested only in the high-energy part of the $Q^2$ term and will arrange the calculation in such a way that it picks up only that part.
    Comparison with prior work will allow us estimate the relative fraction of the high-energy part.

\begin{figure}[htb]
    \centering
    \includegraphics[width=1.0\columnwidth]{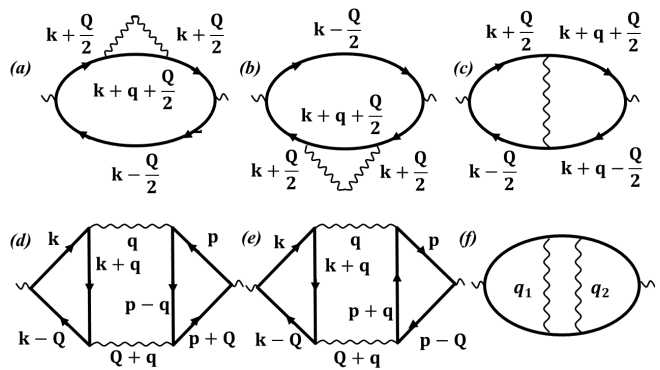}
    \vspace{-0.5in}
    \caption{Diagrams for the
      polarization bubble
    to lowest order in the dynamically screened Coulomb interaction (wavy line). Diagrams {\em a}-{\em c} contribute to the spin susceptibility,
    while diagrams {\em a}-{\em e } contribute to
    the charge susceptibility.
    Diagram {\em f} is a representative
    next-order diagram, which we will use for comparison with the leading order ones.
    }
    \label{fig:diags}
\end{figure}

The minimum set of diagrams for
the renormalized  static polarization bubble, $\tilde\Pi(\bQ)$,
  is shown in Fig.~\ref{fig:diags} {\em a}-{\em e}.
  The interaction correction to the spin susceptibility is given directly by diagrams {\em a}-{\em c}.  [The Aslamazov-Larkin  diagrams {\em d} and {\em e}
 (or the fluctuation diagrams
  in the terminology of Ref.~\onlinecite{ma:1968})
 do not contribute,
 because the traces over Pauli matrices at the vertices vanish in the spin case by SU(2) symmetry.]
 On the other hand, the starting point for the charge susceptibility is the RPA formula $1/\chi_c(\bQ,\Omega)=1/\Pi(\bQ,\Omega)+U_0(\bQ)$, where
 $U_0(\bQ)$ is the bare Coulomb potential. To obtain $\chi_c$ beyond RPA, one simply needs to replace the bare polarization bubble in this formula by the renormalized one, which now includes the contributions of all five diagrams, {\em a}-{\em e}.
 The wavy lines in  Fig.~\ref{fig:diags} correspond to
the Coulomb potential, dynamically screened by free electrons
\bea
U(\bq,\Omega_m)=N_F^{-1}\frac{\kappa_d^{d-1}}{q^{d-1}+\kappa_d^{d-1}\Pi(\bq,\Omega_m)},\label{scr}
\eea
where
$\kappa_2=2\pi e^2 N_F$ and $\kappa_3=\sqrt{4\pi e^2 N_F}$ are the screening momenta in $d=2,3$, correspondingly. To keep the perturbation theory under control, we assume that  $\kappa\ll k_F$ (the high-density approximation) or, equivalently, that
$e^2/v_F\ll 1$. Because typical momentum transfers are expected to be
 of
 order
 $\kappa$,  the polarization bubble
  of free fermions in Eq.~(\ref{scr})
 can be approximated by its small $Q$ limit:
\bea
\Pi(\bq,\Omega_m)&=&1-\frac{|\Omega_m|}{\sqrt{v_F^2q^2+\Omega_m^2}},\label{s_1}\\
\Pi(\bq,\Omega_m)&=&1-\frac{\Omega_m}{v_Fq}\tan^{-1}\frac{v_Fq}{\Omega_m},
\label{s_2}
\eea
in $d=2$ and $d=3$, correspondingly.
 \footnote{By the same argument of small momentum transfers,  the non-analytic, $|\bQ|^{d-1}$ terms in the spin susceptibility, \cite{belitz:1997,chubukov:2003} which
come  from $2k_F$ scattering processes,
appear only at the next order in $\kappa/k_F$ and
 can be neglected.}

If the dynamic interaction is replaced by the static one, $U(\bq,0)$,
 the $Q^2$ term in $\tilde\Pi (\bQ,0)$
 comes
 only
  from low-energy fermions.
  For example, diagram {\em c} in Fig.~\ref{fig:diags} in this case contain a product of two blocks
\bea
&&
\int d\omega_mG(\bk-\bQ/2,\omega_m)G(\bk+\bQ/2,\omega_m)
\nn\\
&&
\times\int d\omega_m' G(\bk'-\bQ/2,\omega'_m)G(\bk'+\bQ/2,\omega'_m),
\label{static}
\eea
where $\bkp=\bk+\bq$.
Because the interaction is static, integrals over $\omega_m$ and $\omega_m'$ in Eq.~(\ref{static}) are independent, and each of them confines the fermionic momenta
 $\bk$ and $\bk'$ to  narrow regions of width $Q$ near the Fermi surface,
  i.e., only fermions with energies smaller that $v_F Q$ contribute.
  This calculation also shows
 that a low-energy $Q^2$ term
could
 not be obtained by Taylor-expanding Eq.~(\ref{static}) to order $Q^2$. Indeed, such an expansion would generate terms of order
  $\int d\omega_m G^2(\bk,\omega_m)$ or higher, which vanish because the poles of the integrand are in the same half-plane of a complex variable $\omega_m$.
 On the contrary, a high-energy $Q^2$ term can be obtained via a straightforward Taylor expansion. On a technical level, this is the main difference between the
  low- and high energy gradient terms.

The $Q$-dependence of the susceptibility arising from the static Coulomb potential (the Hartree-Fock approximation) was calculated numerically in
Ref.~\onlinecite{geldart:1970} for $d=3$  and Ref.~\onlinecite{maldague:1978} for $d=2$, and by a variational method
for $d=2$
in Ref.~\onlinecite{rajagopal:1977}.
The  low-energy contribution to the $Q^2$ term for the static Coulomb potential in discussed in Sec.~\ref{sec:magn} and Appendix~\ref{sec:static}.
 The
 high-energy contributions
 to
diagrams {\em a}-{\em e}
can be singled out
by subtracting off
 the
 static contributions, which amounts to replacing the Coulomb potential by its dynamic part
\beq
U_{\text{dyn}}(\bq,\Omega_m)=U(\bq,\Omega_m)-U(\bq,0).
\label{dyn}
\eeq
This is the effective interaction that we will be using in the next section.

\subsection{Spin
 channel}
\label{sec:spin2Dcont}
The
 corrections to the polarization bubble in the spin channel, $\delta{\tilde \Pi}_s (\bQ)
 \equiv {\tilde \Pi}_s (\bQ,0)- {\tilde \Pi}_s (\bQ\to 0,0),
 $
  are given by diagrams {\em a}-{\em c} in Fig.~\ref{fig:diags}. The calculations for $d=2$ and $d=3$ are very similar, and we present them in parallel. The self-energy diagrams
   {\em a} and {\em b} are equal. Labeling them as shown in Fig.~\ref{fig:diags}, we obtain for their combined contribution
\bwt
\bea
D_{ab}(\bQ)=D_a+D_b=4\int_{k,q} G^2(\bk+\bQ/2,\omega_m)G(\bk-\bQ/2,\omega_m)G(\bk+\bq+\bQ/2,\omega_m+\Omega_m)U_{\text{dyn}}(\bq,\Omega_m),\label{l2}
\eea
while diagram {\em c} gives
\bea
D_{c}(\bQ)=2\int_{k,q} G(\bk-\bQ/2,\omega_m) G(\bk+\bQ/2,\omega_m) G(\bk+\bq+\bQ/2,\omega_m+\Omega_m) G(\bk+\bq-\bQ/2,\omega_m+\Omega_m) U_{\text{dyn}}(\bq,\Omega_m),\nn\\
\label{l3}
\eea
\ewt
where $\int_k\equiv (2\pi)^{-(d+1)}\int d^dk \int d\omega_m$ and $\int_q\equiv (2\pi)^{-(d+1)}\int d^dq \int d\Omega_m$.  For a quadratic spectrum $\ve_\bk=(k^2-k^2_F)/2m$,
the dispersions in Eqs.~(\ref{l2}) and (\ref{l3}) are expanded as
\bea
\ve_{\bk\pm\bQ/2}&=&\ve_{\bk}\pm \frac {1}{2m}\bk\cdot \bQ+\frac{Q^2}{8m},\nn\\
\ve_{\bk+\bq\pm\bQ/2}&=&\ve_{\bk+\bq}\pm \frac {1}{2m} \left(\bk+\bq\right)\cdot \bQ+\frac{Q^2}{8m}.
\eea
Performing corresponding expansions in the Green's functions and keeping only terms of order $Q^2$, we find that the high-energy $Q^2$ term in
 $\delta{\tilde \Pi}_s (\bQ,0)$
consists of four parts
\bea
\delta{\tilde \Pi}_s (\bQ,0)
&=&D_{ab}(\bQ)+D_c(\bQ)\nn\\
&=&
\delta{\tilde \Pi}_{s,1} + \delta{\tilde \Pi}_{s,2} + \delta{\tilde \Pi}_{s,3} + \delta{\tilde \Pi}_{s,4},
\eea
where
\bwt
\bse
\bea
\delta{\tilde \Pi}_{s,1}
&=&\frac{1}{2m^2}\int_{k,q} (\bk\cdot \bQ)^2\left[G^2_k
G^4
_{k+q}+2G^3_k
G^3
_{k+q}+3G^4_k
G^2_{k+q}
+4G^5_k
G_{k+q}
\right]U_{\text{dyn}}(\bq,\Omega_m),\label{chi1}\\
\delta{\tilde \Pi}_{s,2}&=&\frac{Q^2}{2m}\int_{k,q} \left[G_k^2 G_{k+q}^3+2 G_k^3 G_{k+q}^2 +3 G_k^4 G_{k+q}\right]U_{\text{dyn}}(\bq,\Omega_m),\label{chi2}\\
\delta{\tilde \Pi}_{s,3}
&=&\frac{1}{m^2}\int_{k,q} (\bk\cdot \bQ)(\bq\cdot\bQ)\left[G_k^2 G_{k+q}^4+2 G_k^3 G_{k+q}^3 +G_k^4 G_{k+q}^2\right]U_{\text{dyn}}(\bq,\Omega_m),\label{chi3}\\
\delta{\tilde \Pi}_{s,4}
&=&\frac{1}{2m^2}\int_{k,q} (\bq\cdot\bQ)^2\left[G_k^2 G_{k+q}^4 + 2 G_k^3 G_{k+q}^3\right]U_{\text{dyn}}(\bq,\Omega_m)
\label{chi4}
\eea
\ese
\ewt
with
\beq
G_k\equiv G(\bk,\omega_m)\; \text{and}\;G_{k+q}\equiv G(\bk+\bq,\omega_m+\Omega_m).\label{short}\eeq

It can be readily shown that
\bea
&&\int d\omega_m G^n_kG_{k+q}^m=-\frac{m}{n-1}\int d\omega_m G^{n-1}_k G^{m+1}_{k+q}\nn\\
&&=-\frac{n}{m-1}\int d\omega_m G^{n+1}_k G^{m-1}_{k+q}.\label{ident}
\eea
Applying this identity to  Eqs.~(\ref{chi1}) and (\ref{chi2}),
we find that all terms in square brackets cancel each other, and thus $\delta{\tilde \Pi}_{s,1}
= \delta{\tilde \Pi}_{s,2} =0$
This result can be related to the fact that any two-particle correlation function is a gauge-invariant object and,  as such, can contain the interaction potential only multiplied by factor that vanishes at $q\to 0$.\cite{zala:2001} Such a factor ensures that there is no contribution from a potential which is constant in real space, i.e., a delta-function in the momentum space.
There are no such factors in $
\delta{\tilde \Pi}_{s,1}
$ and $
\delta{\tilde \Pi}_{s,2}
$, and therefore they must vanish.

  The combination of the Green's functions in square brackets in $
  \delta{\tilde \Pi}_{s,3}
  $ does not vanish on applying Eq.~(\ref{ident}). This can be related to the fact the integrand contains a common factor of  $q$, and thus $
  \delta{\tilde \Pi}_{s,3}
  $ does not have to vanish identically. However, it still gives no contribution to leading order in $\kappa/k_F$. Indeed, integrating first over $\omega_m$ and approximating $\ve_{\bk+\bq}=\ve_{\bk}+\bv_F\cdot\bq$ with $\bv_F=v_F \bk/k$, we
  find  that each term in square brackets in Eq.~(\ref{chi3}) yields a combination
  \beq
\int d^dq \int d\ve_\bk(\bq\cdot\bQ)\frac{\text{sgn}(\ve_\bk+\bv_F\cdot \bq)-\text{sgn}\ve_\bk}{(i\Omega_m-\bv_F\cdot\bq)^5},\label{int}
\eeq
which is odd in $\Omega_m$.  Since the potential is even in $\Omega_m$, the integral over $\Omega_m$ vanishes, and thus $
\delta{\tilde \Pi}_{s,3}
=0$ to leading order.

The only term that does not vanish by gauge invariance and is non-zero to leading order is thus $
\delta{\tilde \Pi}_{s,4}$. (In Appendix \ref{app:collect}, we demonstrate another way of
 arriving at the same result
  by first combining diagrams Fig.~\ref{fig:diags} {\em a-c} and then expanding the result to order $Q^2$.)
The calculation of $
\delta{\tilde \Pi}_{s,4}$  is fairly straightforward. The frequency integrals in $
\delta{\tilde \Pi}_{s,4}$ are calculated as
\beq
\int \frac{d\omega_m}{2\pi} G^n_kG^{6-n}_{k+q}
=C_{n} \frac{\text{sgn}(\ve_\bk+\bv_F\cdot \bq)-\text{sgn}\ve_\bk}{2\left(i\Omega_m-\bv_F\cdot \bq\right)^5},\label{int2}
\eeq
where $C_{2}=C_{4}=-4$ and $C_{3}=6$.  Now we replace $d^dk/(2\pi)^d$ by $N_Fd\ve_\bk d\vartheta_d/(2^{d}\pi)$, where $\vartheta_d$ is the solid angle in $d$ dimensions,
  and integrate over $\ve_\bk$. The integral over $\ve_\bk$  is confined by the sign functions in Eq.~(\ref{int2}) to the region $(0,|\bv_F\cdot\bq|)$ and gives a factor of $\bv_F\cdot \bq=v_Fq\cos\theta$.   Averaging over the angle between
$\bv_F$ and $\bQ$ (the direction of which is chosen as a reference) yields
\bea
\label{int3}
&&
\int \frac{d\vartheta_d}{2^{d-1}\pi}
\frac{\cos\theta}{(i\Omega_m-v_Fq\cos\theta)^5}
=\frac{1}{(v_Fq)^5}F_d\left(\frac{\Omega_m}{v_Fq}\right),\nn\\
\eea
where
\bse
\bea
F_2(x)&=&\frac{5|x|}{8}\frac{3-4x^2}{\left(x^2+1\right)^{9/2}},\label{F2}\\
F_3(x)&=&\frac{1}{3}\frac{1-5x^2}{\left(x^2+1\right)^4}.\label{F3}
\eea
\ese
Finally, averaging over the angle between $\bq$ and $\bQ$ gives a factor of $1/d$.
After these manipulations, we obtain
\bwt
\bea
\delta{\tilde \Pi}_{s} (\bQ,0) =
\delta{\tilde \Pi}_{s,4}=\frac{2}{d} \frac{1}{\pi^d}\frac{N_FQ^2}{m^2v_F^4}\int^\infty_0 dq q^{d-3} \int^\infty_0 d\Omega_m U_{\text{dyn}}(\bq,\Omega_m) F_d\left(\frac{\Omega_m}{v_Fq}\right).\nn\\
\label{lll}
\eea
\ewt
Now it is convenient to introduce dimensionless variables $x=\Omega_m/v_Fq$ and $y=q/\kappa$. In $d=2$, the integrals over $x$ and $y$ can be solved analytically:
\bwt
\bea
\delta{\tilde \Pi}_{s} (\bQ,0)&=&\frac{1}{
\pi^2}\frac{Q^2\kappa}{m^2v_F^3} \int^\infty_0 dx F_2(x)  \int^\infty_0 dy \left(\frac{1}{y+1-\frac{x}{\sqrt{x^2+1}}}-\frac{1}{y+1}\right)\label{s2d}\\
&=&-\frac{1}{
\pi^2}\frac{Q^2\kappa}{m^2v_F^3} \int^\infty_0 dx F_2(x)\ln\left(1-\frac{x}{\sqrt{x^2+1}}\right)=\frac{1}{
2\pi}\left(\frac{\pi}{32}-\frac{3}{35}\right)N_F\frac{e^2}{v_F} \left(\frac{Q}{2k_F}\right)^2\approx 1.98\times 10^{-3} N_F\frac{e^2}{v_F} \left(\frac{Q}{2k_F}\right)^2.\nn
\eea
\ewt
In $d=3$, the integral over $y$ is solved analytically but the integral over $x$ needs to be solved numerically, which yields
\bea
\delta{\tilde \Pi}_{s} (\bQ,0)
&=&\frac{16 \times(-0.12)}{3\pi^2}N_F\frac{e^2}{v_F} \left(\frac{Q}{2k_F}\right)^2\nn\\
&&\approx -0.064 N_F\frac{e^2}{v_F} \left(\frac{Q}{2k_F}\right)^2.\label{s3d}
\eea
As we mentioned before, Eqs.~(\ref{s2d}) and (\ref{s3d}) give directly the $Q^2$ term in the spin susceptibility: $\chi^s(\bQ,0)-\chi^s(\bQ\to,0)=\delta\tilde\Pi_s(\bQ)$.

Tracing back our steps, we note that all the internal energy scales are of order of effective plasma frequency $\Omega_p=v_F\kappa$: $|\omega_m|\sim |\ve_\bk|\sim v_Fq\sim |\Omega_m|\sim \Omega_p$.
Therefore, the $Q^2$ terms in Eqs.~(\ref{s2d}) and (\ref{s3d}) are indeed of the high-energy type.

For typical values of $q\sim \kappa$ and $\Omega_m\sim v_Fq\sim \Omega_p$, the dynamically screened potential in Eq.~(\ref{scr}) is
 of
 order of $N_F^{-1}$ and does not depend on the electric charge. Although higher-order diagrams contain higher powers of the interaction, the
 interaction is not small in the dimensionless coupling constant of the theory, $e^2/v_F\sim \kappa/k_F$. This may raise
 a concern about convergence of the perturbation theory. Fortunately, this concern is not legitimate, as
 higher-order diagrams have more integrals over internal energy scales, which do bring additional factors of $e^2/v_F$ compared to lowest-order diagrams {\em a}-{\em c}.
  To see this, one can compare, e.g., diagram {\em c} with
  a
  next-order diagram, e.g., diagram {\em f}. The main contribution to diagram {\em c}
   comes from expanding the fermionic dispersion to order $\bq\cdot\bQ$, and then expanding the corresponding Green's function to second order in this parameter,
    which is equivalent to replacing one of the four Green's functions in this diagram by $G^3$. Consequently,
     the $Q^2$ term from diagram {\em c}
      contains six Green's functions, and by power-counting $G^6\propto \kappa^{-6}$. Integrals over $\omega_m$, $\ve_\bk$, $\Omega_m$, and $\bq$ altogether give a factor of $\kappa^{3+d}$, and another $\kappa^2$ comes from $q^2$ in the expansion of dispersion. Overall, the count is $\kappa^{5+d}/\kappa^6=\kappa^{d-1}\propto e^2$ both in $d=2$ and $d=3$. Now, assuming that the main contribution to diagram {\em f} also comes from terms of order $\bq_{1,2}\cdot\bQ$ in the dispersions, we end up
with eight Green's functions $G^8\propto \kappa^{-8}$. The number of the fermionic variables remains the same but the number of bosonic ones
 is doubled, therefore integrations give $\kappa^{2(d+2)}$. With an extra factor of $\kappa^2$ from
  either $q_1^2$ or $q_2^2$, the overall count of diagram {\em f} is $\kappa^{2d+6}/\kappa^8=\kappa^{2(d-1)}\propto e^4$, which is smaller
than diagram {\em c} by a factor of $e^2$.

\subsection{Charge
 channel}
\label{sec:chic}
In addition to the contribution from diagrams {\em a}-{\em c} in Fig.~\ref{fig:diags}, the
 polarization bubble in the charge channel
 also contains
 the contribution from  Aslamazov-Larkin diagrams {\em d} and {\em e}. It will be shown in this section, however,
that the high-energy $Q^2$ terms from  diagrams {\em d} and {\em e}  cancel each other, so Eqs.~(\ref{s2d}) and (\ref{s3d}) apply to  the charge channel as well.

Labeling the diagrams as shown in Fig.~\ref{fig:diags}, we obtain for the sum of diagrams {\em d} and {\em e}:
\bwt
\bea
D_{de}=D_d+D_e&=& 4\int_{k,p,q} U(\bq,\Omega_m) U(\bq+\bQ,\Omega_m)G(\bk,\omega'_m) G(\bk+\bq,\omega'_m+\Omega_m) G(\bk-\bQ,\omega'_m)G(\bp,\omega_m) \nn\\
&&\times \left[G(\bp-\bq,\omega_m-\Omega_m) G(\bp+\bQ,\omega_m)+G(\bp+\bq,\omega_m+\Omega_m) G(\bp-\bQ,\omega_m)\right].
\eea
\ewt
The dispersions are expanded to order $Q^2$ as  $\ve_{\bk\pm \bQ}=\ve_\bk\pm (\bv_\bk\cdot\bQ)+Q^2/2m$. We will also need to expand the interaction, which depends on the magnitude of $\bq$,  as
\beq
U(\bq+\bQ,\Omega_m)=U(\bq,\Omega_m)+\bQ\cdot\hat\bq \partial_{q} U+\frac 12 Q^2\hat h_\bq U ,
\eeq
where $\hat\bq=\bq/q$, $\hat h_\bq\equiv \sin^2\theta_\bq q^{-1}\partial_q +\cos\theta^2_\bq\partial_q^2$, and $\theta_\bq$ is the angle between $\bq$ and $\bQ$.

It is convenient to split $D_{de}$ into two parts. The first part ($D_{de}^{(1)}$) contains all but the terms arising from the $Q^2/2m$ terms in the expanded dispersions, while the second part
($D_{de}^{(2)}$) contains the
 remaining
  $Q^2/2m$ terms. Collecting all $Q^2$ terms in $D_{de}^{(1)}$, we obtain
\bwt
\bea
D_{de}^{(1)}&=&4\int_{k,p,q}\left\{\left[(\bv_\bk\cdot\bQ)^2 U^2_q G^4_kG_{k+q}-(\bv_\bk\cdot\bQ)(\bQ\cdot \hat\bq)U_q\partial_q U_q   G^3_k G_{k+q} +\frac 12 Q^2 U_q\left(\hat h_\bq U_q\right)G_k^2G_{k+q}\right] J^{(2)}_{pq}\right.\nn\\
&&\left.+\left[(\bv_\bk\cdot\bQ)(\bv_\bp\cdot\bQ)U_q^2G^3_kG_{k+q}- (\bv_\bp\cdot\bQ)(\bQ\cdot \hat\bq)U_q\partial_q U_q G^2_kG_{k+q}\right]J^{(3)}_{pq}
+(\bv_\bp\cdot\bQ)^2 U_q^2G_k^2 G_{k+q} J^{(4)}_{pq}\right\},
\eea
\ewt
where shorthands for $G_l$ with $l=k,p,k+q,p+q$ are the same as in Eq.~(\ref{short}), $U_q\equiv U(\bq,\Omega_m)$, and
\beq
J^{(n)}_{pq}=G_p^n\left[G_{p+q}+(-)^n G_{p-q}\right].
\eeq
Now let's define
\beq
F^{(n)}_q=\int d\ve_\bp \int d\omega_m (G_p)^nG_{p+q},
\eeq
where $q=(\bq,\Omega_m)$.
On expanding $\ve_{\bp+\bq}=\ve_\bp+\bv_\bp\cdot \bq$,  it can be readily seen that $F_{-q}^{(n)}=(-)^{n+1} F_{-q}^{(n)}$. Therefore
\bea\int d\ve_\bp \int d\omega_m J^{(n)}_{pq}=F^{(n)}_q+(-)^n  F^{(n)}_{-q} =0\label{van}
\eea
 for any $n$,
and thus $D_{de}^{(1)}=0$.

Next, we collect $Q^2/2m$ terms and obtain
\bwt
\bea
D_{de}^{(2)}&=&2\frac{Q^2}{m}\int_{k,p,q} U_q^2 \left[G^3_kG_{k+q} J^{(2)}_{pq}+G_k^2G_{k+q} G^3_p\left(G_{p+q}+G_{p-q}\right)\right].\label{de2}
\eea
\ewt
The term with $J^{(2)}_{pq}$ vanishes as before by Eq.~(\ref{van}). In the second term, one needs to go one step farther. Integrating the combination $G^3_p\left(G_{p+q}+G_{p-q}\right)$ over $\omega_m$ and $\ve_\bp$, we obtain (up to an inessential prefactor)
\beq
\frac{\bv_\bp\cdot\bq}{\left(i\Omega_m-\bv_\bp\cdot\bq\right)^3}. \label{van2}
\eeq
Likewise, integration of the product $G_k^2G_{k+q}$ over $\omega'_m$ and $\ve_\bk$ yields
\bea
\frac{\bv_\bk\cdot\bq}{\left(i\Omega_m-\bv_\bk\cdot\bq\right)^2}.\label{van3}
\eea
The product of Eqs.~(\ref{van2}) and (\ref{van3}) is odd under a simultaneous change $\Omega_m\to-\Omega_m$ and $\bq\to-\bq$, whereas $U_q$ is even.
 Hence the integral of the second term in Eq.~(\ref{de2}) vanishes as well, which means that  $D_{de}^{(2)}$=0.
  Therefore, the contribution from the Aslamazov-Larkin diagrams vanishes,
  i.e.,
   the high-energy $Q^2$ term in the
   polarization bubble in the charge channel
is the same as in the spin
channel
\bea
\delta{\tilde \Pi}_{c}
 (\bQ,0) = \delta{\tilde \Pi}_{s}
 (\bQ,0).
\eea

\subsection{2D case, lattice dispersion}

\begin{figure}[htb]
    \centering
    \includegraphics[width=1.0\columnwidth]{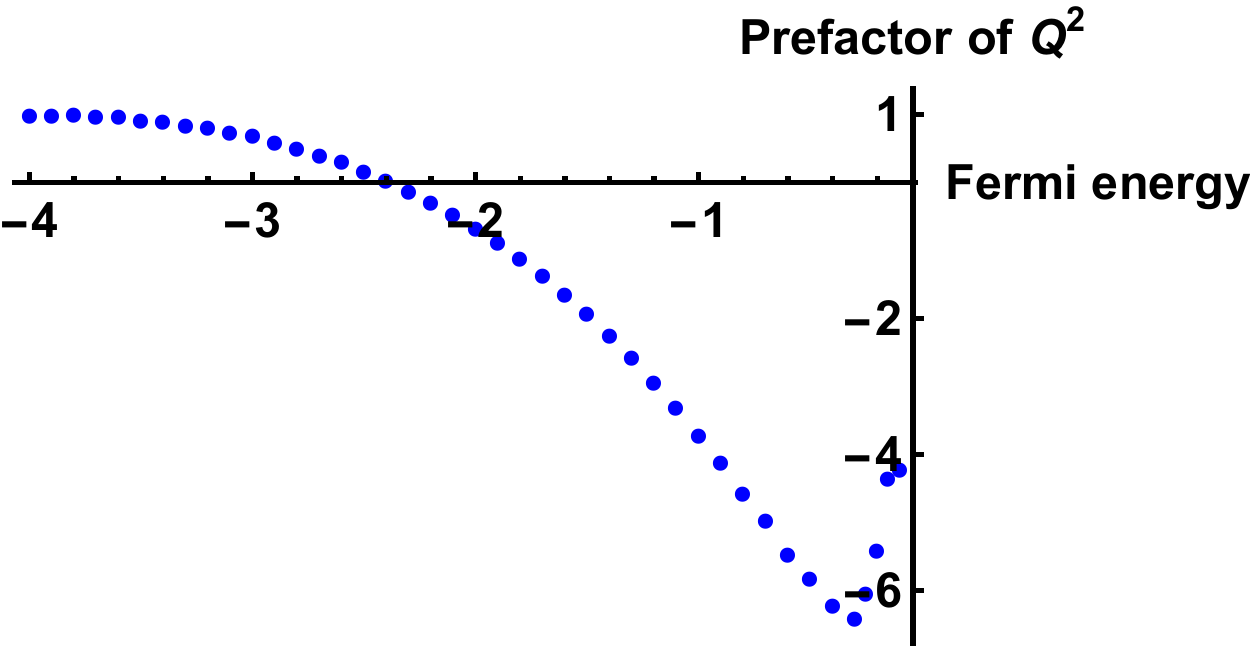}
    \caption{Prefactor of the high-energy $Q^2$ term in the spin susceptibility  for electrons on a two-dimensional square lattice as a function of
    the Fermi energy, measured in units of the hopping integral. $E_F=-4$ corresponds to
     zero filling and $E_F=0$ corresponds to half-filling.
    The numerical result is normalized
     to its value
     at $E_F =-4$. }
    \label{fig:lattice}
\end{figure}

The sign of the high-energy gradient term in
2D
[Eq.~(\ref{s2d})]
 is {\em positive}. If
 the
  low-energy
    $Q^2$ term
    is
     reduced by critical fluctuations, i.e., by
     a factor of
     $m/m^*$,
  then it is the high-energy term that determines the behavior of $\chi^s(\bQ)$ near QCP.
The
positive sign
of this term
would indicate that the susceptibility is peaked at finite $Q$, and
 thus a $Q=0$ QCP is pre-empted by  a finite-$Q$ instability.
 However, the positive sign  was obtained for a special case of fermions with a quadratic spectrum, and its universality needs to be verified. Obviously, the sign will remain the same for any isotropic  but not necessarily quadratic dispersion: in this case one just needs to replace a factor of $1/m$ in Eq.~(\ref{chi4}) by the inverse effective mass $\partial^2_k
\ve_\bk$. To check whether  the sign can be reversed in the presence of a lattice, we computed numerically the prefactor of the high-energy gradient term in the same
 model with a long-range Coulomb interaction but for electrons on a square lattice  with a tight-binding dispersion
 $\ve_\bk=-2t
 (\cos k_x
 +\cos k_y)$.
 \footnote{Strictly speaking, the interaction of electrons on a lattice must be described by matrix elements of the bare Coulomb potential in the basis of Wannier functions. This changes the magnitude of the Coulomb potential but not its $1/q^{d-1}$ form. As we are interested here only in the sign of the effect, we ignore this complication and use the same bare Coulomb potential as for electrons in continuum.}  The dynamically screened Coulomb interaction is the same as in Eq.~(\ref{scr}), except for now $\Pi(\bq,\Omega_m)$  is the polarization bubble of electrons on a square lattice, which we computed numerically for
a range of filling factors.
The calculation follows the same lines as in Sec.~\ref{sec:spin2Dcont}. We focus on the leading contribution to the $Q^2$ term [$
\delta{\tilde \Pi}_{s,4}$ in Eq.~(\ref{chi4})],
 calculate the integrals over $\omega_m$ and $\ve_\bk$ analytically, and compute the remaining four-dimensional integral
  over $\bq$, $\Omega_m$, and along the Fermi surface
   numerically. By $C_4$ symmetry, the $Q^2$ term is isotropic. In Fig.~\ref{fig:lattice}, we plot its prefactor as a function of the Fermi energy,
   measured in units of $t$, such that $E_F=-4$ corresponds to the bottom of the band.
   For $E_F\to -4$, the numerical solution reproduces the analytic result in Eq.~(\ref{s2d}) with high accuracy. However, at larger fillings, we found a new behavior.
    Namely, Fig.~\ref{fig:lattice} shows that the prefactor of the $Q^2$ term changes sign around quarter-filling
    ($E_F =-2$)
    and remains negative  all the way up to half-filling
    ($E_F =0$).
     A negative prefactor for $Q^2$
implies that
 $\chi ^s(\bQ)$ has a maximum at $\bQ=0$, as expected near QCP towards an order with $\bQ=0$.  We see therefore that lattice
tends to stabilize a $Q=0$ QCP,  given that the maximum of $\chi^s(\bQ)$ at $\bQ=0$ is higher than that at finite $\bQ$.

\subsection{Relative magnitudes of high-energy and low-energy contributions to
the $Q^2$ term}
\label{sec:magn}

We found in previous
sections that the magnitude of the prefactor
of the high-energy $Q^2$ term,
 in units of $e^2/v_F$,
 happens
 to be numerically small both in 2D and 3D.
 This naturally
 raises
 a
 question
 about
 the ratio
 of
 high-energy
 to
  low-energy
  parts
 of the $Q^2$ term.

In 3D, the low-energy contribution is non-zero already for free fermions, whereas the high-energy contribution comes
only from interaction and is therefore parametrically small
at small $r_s$.  The interaction correction to the low-energy part is small as well.
 For $r_s\sim e^2/v_F\sim 1$, all
   contributions to $Q^2$
    become of the same order, and
    it makes sense
 to compare the numerical prefactors.
The same set of diagrams as in Fig.~\ref{fig:diags} for the
  polarization bubble
 was computed
 numerically in Ref.~\onlinecite{khalil:2002} for $r_s =2$.
  We digitized the data of Ref.~\onlinecite{khalil:2002},  fitted the small-$Q$ parts of the curves by a $Q^2$ form, and
   extracted the total interaction-dependent
  prefactor of  the $Q^2$ term, which contains both low- and high-energy contributions.
  Comparing the total numerical prefactor with the analytic result for the high-energy contribution, given by Eq.~(\ref{s3d}),
  we found that the high-energy contribution
  amounts to about $15\%$ and $30\%$ of the total for the charge and spin susceptibilities, correspondingly.
   \footnote{Reference \onlinecite{khalil:2002} gives the results for diagrams {\em a}-{\em c} and {\em d-e} separately, from which one can extract 
   the polarization bubbles both in
   the charge and spin channels.}   At the same time, the total interaction-induced $Q^2$ term in the spin and charge susceptibilities for $r_s=2$ is about $70\%$
    and $140\%$ of the free-fermion result, correspondingly.

 The 2D case
 is
 different in that  the free-fermion polarization bubble for fermions with a parabolic dispersion
 is independent of $Q$ up to $Q = 2k_F$,
  i.e., the prefactor
  of the
  $Q^2$ term
  is zero.
  This
   degeneracy
   can be
    broken by
   lattice
   but,
   at low enough filling,
   the
   $Q^2$ term in
   the
   free-fermion bubble is
   still
   small. Then both high-
   and low-energy contributions to the $Q^2$ term come from interaction,
   which allows for a direct comparison between the two contributions.
   In what follows we focus on the spin susceptibility.
   We estimated  the
     low-energy contribution to the interaction-induced $Q^2$ term by
   evaluating the diagrams in Fig.~\ref{fig:diags}(a-c)
    with a
     statically screened Coulomb potential, $U(\bq,0)=2\pi e^2/(q+\kappa)$.
   (In Sec.~\ref{sec:spin2Dcont} we subtracted off this contribution.)
 We obtained  (see
 Appendix \ref{sec:static} for details):
  \bea
 \delta{\tilde \Pi}^{\text{low}}_{s}
 (\bQ)
 &=&\frac{1}{12\pi}
 \left( \frac{92}{15}-\ln{\frac{2\sqrt{2}}{r_s}}\right) N_F \frac{e^2}{v_F} \left(\frac{Q}{2k_F}\right)^2\nn\\
 &=&\left(0.16-0.027\times\ln \frac{2\sqrt{2}}{r_s}\right)N_F \frac{e^2}{v_F} \left(\frac{Q}{2k_F}\right)^2,\nn\\
\label{low}
  \eea
  where $r_s=\sqrt{2} e^2/v_F=2\sqrt{2}\kappa/k_F$ in 2D.

   The $Q$-dependence of the polarization bubble in 2D due to static Coulomb interaction was addressed  in Refs.~\onlinecite{rajagopal:1977} and \onlinecite{maldague:1978}.
   There is some confusion about the results in prior literature which needs to be clarified. Namely, the numerical calculation in Ref.~\onlinecite{maldague:1978} was
    performed for
    the
    bare Coulomb potential ($\kappa=0$)
     and produced a finite result for the susceptibility at all $Q$,
      whereas in Ref.~\onlinecite{rajagopal:1977} it was argued that the prefactor of the $Q^2$ term is divergent at $\kappa\to 0$. Actually, 
      these two results do not contradict each other.
    We found,
    in agreement with Ref.~\onlinecite{rajagopal:1977},
     that the leading term in the $Q$ dependence of $\delta{\tilde \Pi}^{\text{low}}_{s}
 (\bQ)
$ for the bare Coulomb potential is $Q^2\ln Q$ rather
     than $Q^2$. If screening is taken into account, $\ln Q$ is replaced by $\ln \kappa/k_F\sim \ln r_s$, as it is the case in Eq.~(\ref{low}).  At the same time,
     we see that the numerical prefactor of the $Q^2\ln r_s$ term
     in Eq.~(\ref{low})
      is much smaller (by a factor of $\approx 6$) than that of the $Q^2$ term, so the numerical
     results of Ref.~\onlinecite{maldague:1978} are indeed well-described by the $Q^2$ form except for the region of  extremely small $Q$.

 The $Q$-dependence of the polarization bubble in 2D due to the dynamically screened Coulomb interaction was also calculated numerically in Ref.~\onlinecite{khalil:2002}.
 Fitting the result of Ref.~\onlinecite{khalil:2002} for the spin susceptibility into a $Q^2$ form at $r_s=2$, we found that it is described by Eq.~(\ref{low}) very well.
 This implies that
 the interaction-induced $Q^2$ term comes primarily from low energies.
 Comparing the numerical result for the total interaction-induced $Q^2$ term with our analytic result for its high-energy part [Eq.~(\ref{s2d})],
 we found that the latter amounts to about $2\%$ of the former. (For the charge channel, the fraction of  the high-energy part is about $4\%$.)

Another comment on the 2D case is in order. That the susceptibility of
 non-interacting electrons with  quadratic dispersion is flat up to $2k_F$ indicates a high degree of frustration
   with respect to ordering into a state with finite  $Q$.
   (The same flatness holds for 2D fermions with a
 Dirac  dispersion. \cite{kotov:2012})
  The
 Coulomb interaction lifts
 this
  degeneracy, and
   numerical calculations show that the
    full
    susceptibility is peaked at  $Q=2k_F$.\cite{maldague:1978,khalil:2002} (Note in this regard that the prefactors
 of
 the $Q^2$ terms in Eqs. (\ref{s2d}) and (\ref{low})
 are positive, i.e., the susceptibility increases with increasing $Q$.)    For electrons on lattice,
the tendency to ordering at finite rather than zero $Q$ is pronounced already
in the non-interacting case: numerical calculations for square, triangular, and honeycomb lattices show that the
 susceptibility is peaked at the momenta connecting certain points on the FS.\cite{batista:2016,batista:2016b,batista_private}
   In light of these results, a 2D electron system may not seem to be a good candidate for a $Q=0$ instability. Nevertheless,
    the type of an instability in an interacting system is decided by the relative strengths of renormalizations of $\chi^s$ at $Q=0$ and finite $Q$.
     The former is determined by FL parameters, while the latter cannot be quantified in this way and depends on details of the electron spectrum and interaction.
      Even though $\chi^s(\bQ)$ of free electrons may have a peak at finite $Q$,
      the $Q=0$ peak in
      renormalized $\chi^s (\bQ)$
       may be  higher,
       and thus the $Q=0$ instability may win. In addition, the Kohn anomaly, which leads to  a peak at $2k_F$, is weakened under certain circumstances,
        e.g., in chiral electron systems,  such as graphene and surface states of 3D topological insulators, due to suppression of
        backscattering into states with opposite (pseudo) spins.\cite{kotov:2012}

\section{
A
$Q^2$ term in the Hubbard model}
\label{sec:Hubbard}

As another example, we compute the polarization bubble of interacting fermions assuming that the four-fermion interaction is
 Hubbard-like, i.e., it is equal to a constant $U$ for $q$ below some cutoff,
 $q_c$,
   and is zero for larger $q$. To simplify calculations, we set $q_c$ to be much smaller tan $k_F$.
   This will allow us to use a  small-$q$ and small-$\Omega_m$ form of the polarization bubble, Eq. (\ref{s_1}).

\begin{figure}[htb]
    \centering
    \includegraphics[width=1.0\columnwidth]{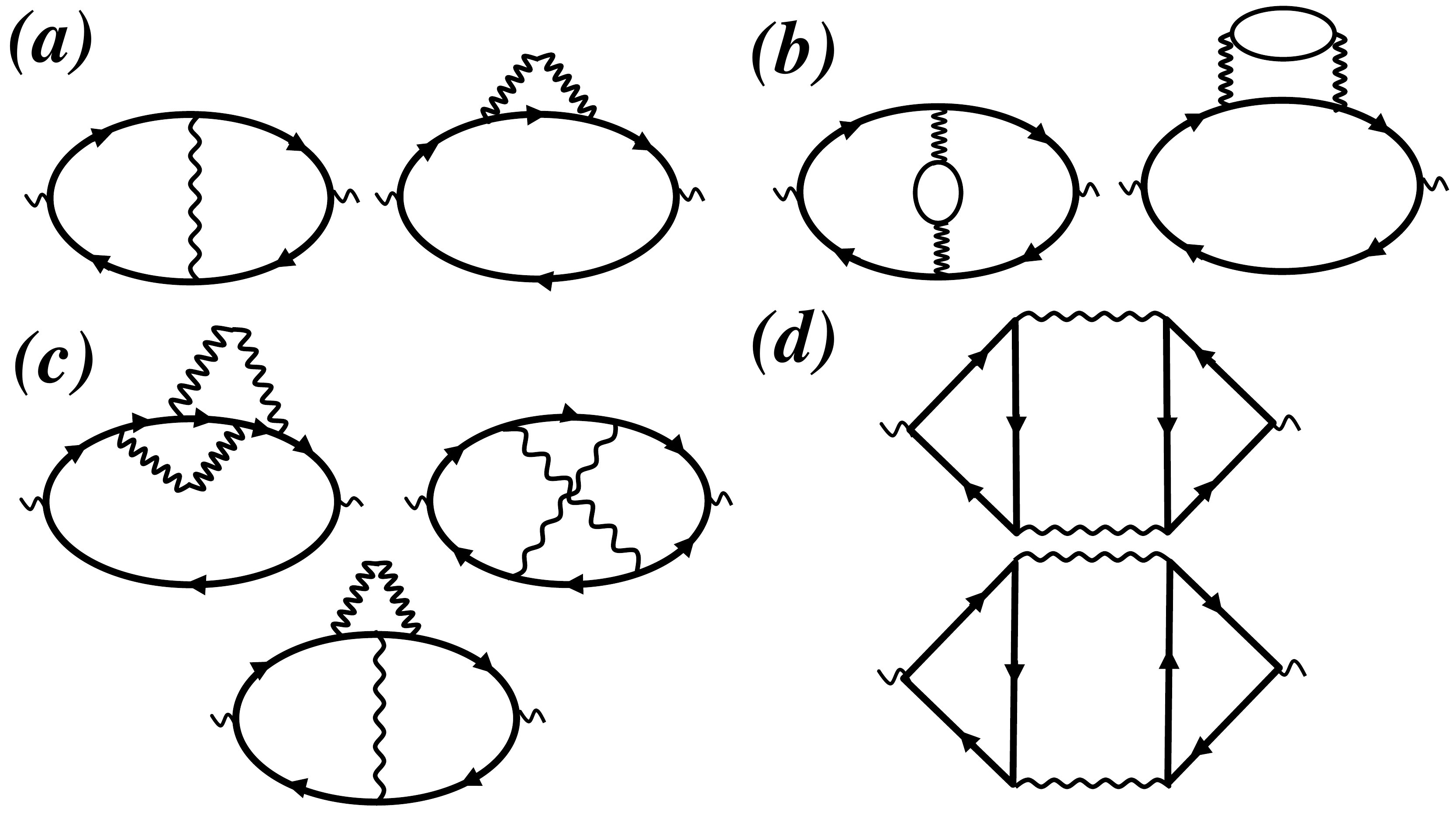}
    \caption{Diagrams for the polarization bubble in the Hubbard model to second order in $U$.  }
    \label{fig:hubbard}
\end{figure}

To first order in $U$ (panel {\em a} in Fig.~\ref{fig:hubbard}), the self-energy diagram amounts to shifting the chemical potential, while the vertex diagram is reduced
to a product of two free-fermion polarization bubbles. None of the above produces a
high-energy $Q^2$ term.
However, the situation changes at
 second order in $U$ because now self-energy and vertex renormalizations within a particle-hole bubble can be viewed as
 dynamical screening of the interaction between high-energy fermions.  In this respect, the
 Hubbard
 model, taken at order $U^2$, becomes similar to the  model with  dynamically screened Coulomb interaction, taken at
 first order in this interaction.  It remains to be seen, however, whether the prefactor
 of the $Q^2$ term in the Hubbard model is of the same sign and comparable magnitude as for the model with Coulomb interaction.
  Interaction-induced corrections to
  the
   static polarization bubble in the Hubbard model
  were
  analyzed in
  Ref.~\onlinecite{chubukov:1993} without
  making a
  distinction between low-energy and high-energy contributions. Our goal is to distinguish between the two.

  There are seven
  distinct non-RPA diagrams for
  renormalization of $\Pi$ to second order in $U$,
   which potentially can give rise to a high-energy $Q^2$ term.
   We show them
  in
    panels {\em b}-{\em d} of
    Fig.~\ref{fig:hubbard}.
   Five of these diagrams, shown in panels {\em b} and {\em c},
    renormalize
  both the charge and spin susceptibilities,
  while the
  Aslamazov-Larkin
  diagrams, shown in panel {\em d},
  renormalize only the charge susceptibility.

  For definiteness, we consider 2D case and approximate the fermionic dispersion by a parabolic one.
 We found that  
 the three diagrams in panel {\em c}
  and  two Aslamazov-Larkin diagrams  in panel {\em d}
 are smaller in $q_c/k_F$ than the two diagrams in panel {\em b}.
   The
    two
    last
     diagrams
    can be viewed as
   vertex and self-energy corrections to the particle-hole bubble
    coming from
    effective interaction
    $U_{\text{eff}} (q, \Omega_m) =
   -U^2 \Pi (q, \Omega_m)$.
      To explore the region of somewhat larger $U$, we will extend this formula to the full RPA result
     $U_{\text{eff}} (q, \Omega_m) = U
     \left[1 +
   U\Pi (q, \Omega_m)\right]^{-1}$.
    The actual diagrams that need to be evaluated are then the first-order diagrams (panel {\em a}) in which  $U$ is replaced by $U_{\text{eff}}$.
  The second-order result can be obtained by expanding the effective interaction back to second order in $U$.
   Expanding these diagrams
  to
order $Q^2$,
and
integrating over fermionic  frequencies and momenta in the same way as in
Sec.~\ref{sec:coulomb}, we obtain
for
the
high-energy $Q^2$ contribution to the
polarization bubble in the spin channel
\beq
\delta {\tilde\Pi}_s (\bQ,0) =  N_F A_H\frac{Q^2}{k^2_F},
\label{n_3}
\eeq
where
\beq
A_H = \frac{2 k^2_F }{m^2 v^3_F} I_H
\label{n_4}
\eeq
and
\bea
I_H &=&
 \int \frac{dq}{2\pi} \int_{-\infty}^{\infty} \frac{dx}{2\pi}
\left[F_2(x)- \frac 14\delta(x)\right]
\frac{U}{1 +
UN_F
 \left(1- \sqrt{\frac{x^2}{x^2+1}}\right)}.\nn\\
\label{n_5}
\eea
Here,
 $x = \Omega_m/
 v_F q $ and $F_2(x)$ is given by Eq.~(\ref{F2}).
  One can check that the remaining integral over $x$
  vanishes if
  the effective interaction
  is approximated
  as static, i.e.,
  the last factor in the integrand of  Eq.~(\ref{n_5})
  is replaced
  by
  $U/(1+ U N_F)$.
   We follow the same strategy as in
    Sec.~\ref{sec:coulomb} and
     just subtract off the static interaction from Eq.~(\ref{n_5}).  Then the delta-function term
 in Eq.~(\ref{n_5})
 can be omitted, and the expression for $I_H$ becomes
  \bwt
  \beq
  I_H = -\frac{5}{8\pi}  \int \frac{dq}{2\pi}\int_0^{\infty} \frac{dx x (4x^2-3)}{(x^2+1)^{9/2}}\left[ \frac{U}{1 +
 UN_F
  \left(1- \sqrt{\frac{x^2}{x^2+1}}\right)} - \frac{U}{1 +
  U N_F
  }\right].
 \label{n_7}
 \eeq
\ewt
This expression is similar to
the corresponding formula for  Coulomb interaction in 2D [Eq.~(\ref{lll})] with one important distinction.  For the Coulomb case,
the interaction behaves as $1/q$ and,
as
a consequence, the integral over $q$ in Eq.~(\ref{lll})
is ultraviolet convergent, i.e.,
integration over $q$ can be extended to infinity.
  For the Hubbard case, the
  integral over $q$ in  Eq.~(\ref{n_7}) is not convergent and needs to be cut off at $q_c$.

\begin{figure}[htb]
    \centering
    \includegraphics[width=1.0\columnwidth]{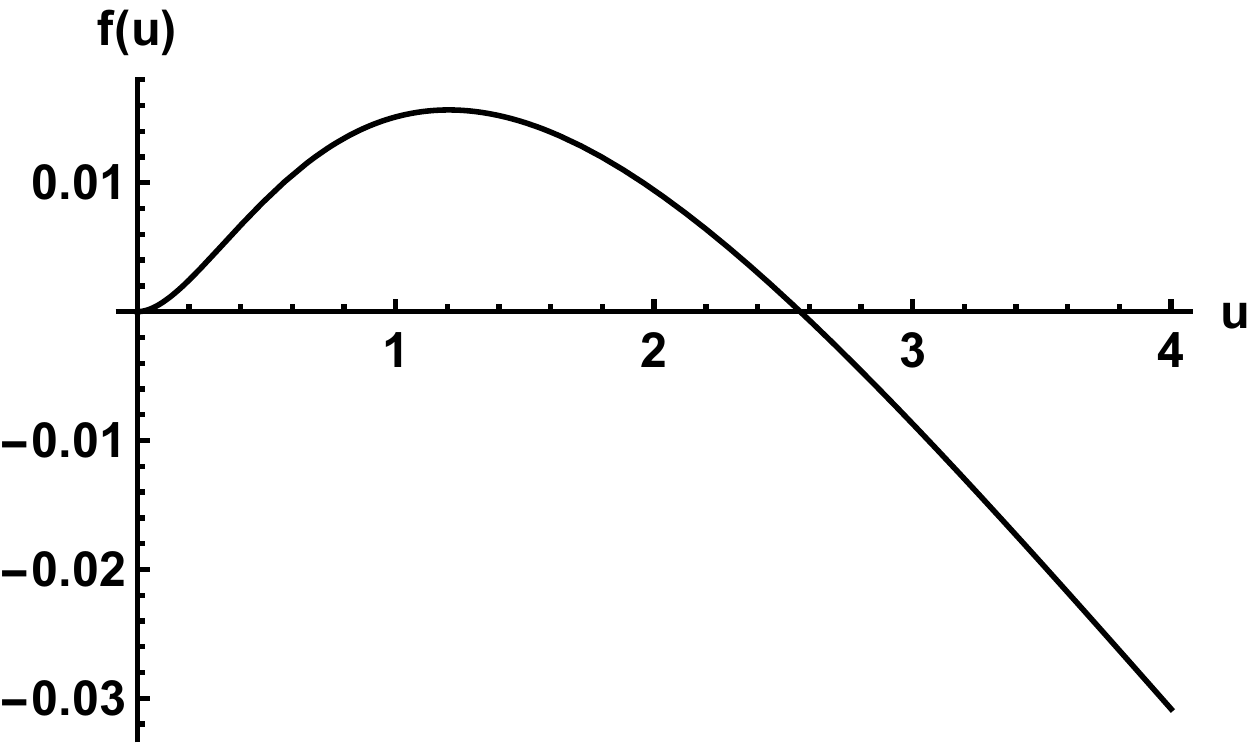}
    \caption{Function $f(u)$ [Eq.~(\ref{fu})].}
    \label{fig:fu}
\end{figure}
Substituting Eq.~(\ref{n_7}) into Eq.~(\ref{n_4}), we obtain
\beq
\delta {\tilde \Pi}_s (\bQ,0) = N_F
\left(\frac{q_c}{k_F}\right)~\left(\frac{Q}{k_F}\right)^2~f(U N_F
),
\eeq
where
\bea
f(u) &=& - \frac{5}{2} u\int_0^{\infty} \frac{dx x (4x^2-3)}{(x^2+1)^{9/2}}\nn\\
&&\times\left[ \frac{1}{1 + u \left(1- \sqrt{\frac{x^2}{x^2+1}}\right)} - \frac{1}{1 + u}\right].\label{fu}
\eea
At small $u$,
\beq
f(u) =  \frac{277 \pi }{16384 \sqrt{2}}u^2 + O(u^3) \approx 0.075 u^2
\eeq
We see
that, at weak coupling,
 the prefactor
of
 the $Q^2$ term in $\delta {\tilde \Pi}_s (\bQ,0)$ is positive.
  As the consequence,  $\chi^s (\bQ)$ imcreases with increasing $Q$. This is similar to what we obtained for the Coulomb interaction in 2D for a parabolic dispersion. Also,
 as
 in the Coulomb case, the magnitude of the high-energy contribution to the $Q^2$ term is 
 numerically
 very small.

Function $f(u)$ in plotted Fig.~\ref{fig:fu}.  We see that the magnitude of $f$ remains small for all $u$, but $f(u)$ changes sign at $u 
\approx
2.6$ and becomes negative for
larger $u$. At such $u$, $\chi^s (\bQ)$ becomes a decreasing function of $\bQ$,
 i.e., $\chi^s (\bQ)$ has
 at least a
  local maximum
 at $\bQ=0$.

We note, however, that the
 prefactor
 of  the  $Q^2$ term  changes its sign
 at a rather large
 $u$. Whether at such $u$ the system still does not order magnetically is unclear (within RPA, the Stoner instability occurs at $u_{cr}=1$, but the value of $u_{cr}$
 changes due to corrections beyond RPA).

A low-energy contribution to the $Q^2$ term in $
\delta {\tilde \Pi}_s (\bQ,0)$ in a 2D system with a parabolic dispersion also emerges to second order in $U$.
A computation of this 
contribution
is rather involved
 and we 
 left it aside.
We note, however, that the low-energy contribution is non-zero already for free fermions, once we include higher powers of $k^2$ into an isotropic dispersion
  or put
  the model
  on a lattice.

\section{Discussion and conclusions}
\label{sec:analysis}
We now discuss the results of the previous sections  in the context of a quantum-critical theory. We have shown that there are two contributions to the gradient ($Q^2$)  term in the bosonic propagator near a $Q=0$ QCP. One comes from fermions with high energies, by which we understand energies
above the upper cutoff
of
 the effective
 boson-fermion low-energy theory, $\Lambda$.
Another comes from fermions with low energies, of order of $v_F Q$. The  low-energy contribution is present already in the bosonic susceptibility made  of free fermions
 (except for the special cases of, e.g., linear and  parabolic dispersions in 2D). To get a high-energy contribution to the
$Q^2$ term, one has to include dynamical screening of the interaction between high-energy fermions.
If only static screening is included, the high-energy contribution is absent.

The low-energy part of the $Q^2$ term
near a $Q=0$ QCP is not reduced by
a
small ratio
of
 $\Lambda$
 to
the  fermionic bandwidth.
However, it is reduced near QCP by
a divergence
in the effective mass $m^*$.  There is no
general formula relating the prefactor of the low-energy
$Q^2$ term
to the Landau parameters. However,
if near QCP the fermionic self-energy depends on the frequency much stronger than  on the
momentum,
the low-energy part of the $Q^2$ term is reduced by
a factor of
$m/m^*
=\left[1
- \partial \Sigma (\omega)/\partial\omega\right]^{-1}
$.  The high-energy contribution to $Q^2$ term is not reduced by $m/m^*$ and, in general, has to be the dominant contribution
near QCP.

 We found, however, that, at least in
two microscopic models,
the high-energy contribution to the $Q^2$ term is numerically very small.
The most spectacular example is a 2D Fermi gas with Coulomb interaction -- the high-energy contribution to $Q^2$ term in the bosonic susceptibility is less than
two percent of the total.

Based on these numbers, one can envisage two possible types of quantum-critical theories. In the first one, adopted in earlier studies, the numerical
smallness of the high-energy terms is disregarded as an artifact of a particular  model. The starting point
 for such a theory is a high-energy action with a $Q^2$ term, which is not assumed to be small.
Such a theory
 has dynamical exponent $z=3$ and
 yields the
 fermionic self-energy $\Sigma (\omega) \propto \omega^{d/3}$,
   modulo logarithmic corrections.\cite{metlitski:2010b,metlitski:2010c}
 In the second type of theories, the smallness of the high-energy $Q^2$ term is treated as the real
effect,
and the bosonic propagator at energies ${\cal O}(\Lambda)$ is taken to be independent
of
 $\bQ$, at least to first approximation.  Then the
entire $Q^2$
term
in
 the bosonic propagator comes from low energies, and
 its
  prefactor depends on $m/m^*$. Because in critical
  boson-fermion
   theories the
   fermionic self-energy, and hence
   $m/m^*$, depends on
   the prefactor of the $Q^2$ term,
   one now has to solve for $m/m^*$ self-consistently, keeping $m/m^*$ in the prefactor
   of the
   $Q^2$ term.  Such a procedure has not been yet implemented for the critical behavior near a $Q =0$  QCP.

The self-consistent procedure
 of this kind
 was implemented
in Refs.~\onlinecite{wolfle:2011,abrahams:2014}
 for an antiferromagnetic QCP (an instability at $\bQ = \bQ_0$).
In this theory,
the prefactor
of
the $(\bQ - \bQ_0)^2$ term in the bosonic propagator is set phenomenologically to scale as  $(m/m^*)^2$, and
the consequences
of this choice
for the quantum-critical behavior are analyzed.  At
 finite $\bQ_0$, the high-energy contribution to the gradient term is non-zero already for free fermions, and in general it is not small.
 At the same time, the low-energy contribution is reduced by a factor of $\Lambda/W$.  Then, in a generic case,
  the largest contribution to the gradient term should come from high energies,
as it is assumed in conventional boson-fermion theories near a $Q=0$ QCP.\cite{abanov:2003}
 However, in light of our results for the $Q=0$ case, it
 would  be interesting to analyze the high-energy
$Q^2$ terms
in a broader set of models.

\acknowledgements
We thank N. Prokofiev for his help with the numerical calculation, and  E. Abrahams, C. Batista, S. Maiti, P. Wolfle, and V. A. Zyuzin for stimulating discussions.
P.S. acknowledges support from the Institute for Fundamental Theory, University of  Florida.
The work of A.V.C. was supported by the NSF via grant DMR-1523036.

\newpage

\bwt
\appendix
\section{Diamagnetic susceptibility as a low-energy property}
\label{sec:dia}

It is commonly assumed that the diamagnetic susceptibility of a free Fermi gas, $\chi^{\text{dia}}$, is determined by all occupied states, both far below and near the Fermi energy.\cite{levitov,tsvelik} In our terminology, this makes $\chi^{\text{dia}}$ a high-energy property. This argument is based on the thermodynamic way
of calculating $\chi^{\text{dia}}$,\cite{statphysI} in whihc one first finds the Landau levels, then calculates the free energy as a sum over these levels, and finally differentiates the result over  the magnetic field. Such a procedure can be carried out only in those cases
when the exact form of the Landau spectrum is known, e.g., for parabolic and linear dispersions. However, one can also calculate $\chi^{\text{dia}}$ within the linear-response theory,\cite{hebborn:1970,levitov,tsvelik} as a prefactor of the $Q^2$ term in the current-current correlation function.
The result of such a calculation, Eq.~(\ref{dia}), which can be carried out for an arbitrary electron dispersion, is similar to that for the prefactor of the $Q^2$ term in the charge or spin susceptibilities [Eq.~(\ref{A})] in that $\chi^{\text{dia}}$ is entirely parameterized by the derivatives of the dispersion at the Fermi energy. This implies that  $\chi^{\text{dia}}$ is, in a fact, a low-energy property. Some details of the calculation are presented below.

As in Ref.~\onlinecite{maslov:2009}, we assume that the electron dispersion is an isotropic but otherwise arbitrary function of $\bk$. With $\bQ$ chosen as the $z$-axis, we need to calculate the $Q^2$ term in the $xx$ component of the current-current correlation function
\beq
K_{xx}=-2\frac{e^2}{c}T\sum_{\omega_m}\int \frac{d^{3}k}{(2\pi)^3} v^{x}_{\bk+Q\hat z}v^x_\bk G(\bk+Q\hat z,\omega_m) G(\bk,\omega_m),
\eeq
where $v^{x}_\bk=\partial\ve_{\bk}/\partial k_x$. Defining
$\delta K_{xx}(Q)\equiv K_{xx}(Q)-K_{xx}(0)$, the diamagnetic susceptibility is found as $\chi^{\text{dia}}=\lim_{Q\to 0}\delta K_{xx}(Q)/cQ^2$.
Summation over $\omega_m$ gives
\beq
K_{xx}=-2\frac{e^2}{c}\int \frac{d^{3}k}{(2\pi)^3} v^2_{x} \frac{n_F(\ve_{\bk+Q\hat z})- n_F(\ve_\bk)}{\ve_{\bk+Q\hat z}-\ve_\bk},
\eeq
where $n_F(E)$ is the Fermi function.
Next, we expand the dispersion, the $x$-component of the velocity, and the Fermi function to order $Q^2$ as
\bea
\delta\epsilon&\equiv& \ve_{\bk+Q\hat z}-\ve_{\bk} =v_k Q  \cos\theta +\frac 12 Q^2\left(\sin^2\theta\frac{v_k}{k}+\frac{1}{m_k}\cos^2\theta\right),\nn\\
v^x_{\bk+Q\hat z}&=&\sin\theta\cos\phi\left[v_k+Q\cos\theta\left(\frac{1}{m_k}-\frac{v_k}{k}\right)+\frac{Q^2}{2k}\left\{\left(3\cos^2\theta-1\right)\left(\frac{v_k}{k}-\frac{1}{m_k}\right)+k\gamma_k\right\}\right],\nn\\
&& \frac{1}{\delta \epsilon}\left[n_F(\ve_\bk+Q\hat z)- n_F(\ve(\bk)\right] = n'_F(\epsilon_k)+\frac 12 n''_F(\epsilon_k)\delta\epsilon+ \frac 16  n'''_F(\epsilon_k)\delta\epsilon^2.
\eea
Here,  $(\theta,\phi)$  are the polar and azimuthal angles of $\bk$, $\epsilon_k\equiv \ve_{\bk}+E_F$,
$v_k=d\epsilon_k/dk$ is the group velocity, $1/m_k=d^2\epsilon_k/dk^2$ is the effective mass, and $\gamma_k=d^3\epsilon_k/dk^3$. For a parabolic spectrum, $v_k=k/m$, $m_k=m=\text{const}$, $\gamma_k=0$, and $v^x_{\bk+Q\hat z}$ is independent of $Q$.
At $T=0$, the derivatives of the Fermi functions are replaced by $ d^{p}n_F/d\epsilon_k^p= -\delta^{(p-1)}(\epsilon_k-E_F)$. Subtracting off the $Q$-independent term from $K_{xx}$, integrating  over $\epsilon_k$ by parts, and averaging over the angles, we arrive at the final result for the diamagnetic susceptibility:
\bea
\chi^{\text{dia}}= \frac{e^2}{15 c^2} \left\{ N(\epsilon_k) \left(  \frac{
v_k}{ k m_k}  - \frac{v_k^2}{ k^2}
 +\frac 12 v_k \gamma_k
\right)  + \frac{\partial}{\partial \ve_k}\left[ -  \frac{N(\epsilon_k)}{4} \left( \frac{2 v_k^3 }{ k } + \frac{3 v_k^2}{ m_k} \right) + \frac{1}{6}\frac{\partial}{\partial \epsilon_k} \left(N(\epsilon_k) v_k ^4  \right)\right]\right\}\Big\vert_{\epsilon_k = E_F}.\label{dia}
\eea
For a parabolic dispersion,  the last equation is reduced  to
\beq
\chi^{\text{dia}}= -\frac{e^2 N_F}{12 m^2c^2} = -\frac{1}{3} \chi^s,
\eeq
as it should.

\section{A manifestly gauge-invariant way of collecting diagrams}
\label{app:collect}
\begin{figure}[htb]
    \vspace{-2in}
    \centering
    \includegraphics[width=1.0\columnwidth]{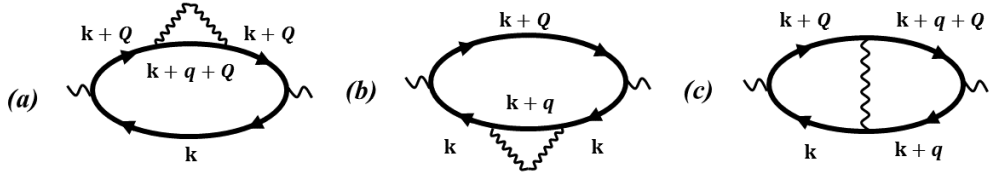}
    \vspace{-2in}
    \caption{Diagrams for the polarization bubble to first oder in the statically screened Coulomb potential. }
    \label{fig:diags2}
\end{figure}

In this Appendix, we show how diagrams {\em a}-{\em c} in Fig.~\ref{fig:diags} can be combined in a manifestly gauge-invariant way. For convenience, we relabel the diagrams as shown in Fig.~\ref{fig:diags2} and adopt ``relativistic'' notations: $k=(ik_0,\bk)$, $q=(iq_0,\bq)$, and $Q=(iQ_0,\bQ)$.  $Q_0$ will be set to zero later on in the calculation.

The sum of the self-energy diagrams ({\em a} and {\em b}) can be written as
\bea
D_{ab}=-2 \int_k G_kG_{k+Q}\left(G_{k+Q}\Sigma_{k+Q}+G_k\Sigma_k\right),
\eea
where
\beq\Sigma_k=-\int_q U_qG_{k+q}
\label{self}
\eeq
 is the one-loop self-energy.

Re-writing the product of the Green's functions as
\bea
G_kG_{k+Q}&&=\frac{1}{iQ_0-\ve_{\bk+\bQ}+\ve_\bk}\left[G_k-G_{k+Q}\right],\label{split}
\eea
we represent $D_{ab}$ as a sum of two parts $D_{ab}=D_{ab}^{(1)}+D_{ab}^{(2)}$,
where
\bea
D_{ab}^{(1)}&=&2\int_k \frac{G_kG_{k+Q}\left(\Sigma_k-\Sigma_{k+Q}\right)}{iQ_0-\ve_{\bk+\bQ}+\ve_\bk},\nn\\
D_{ab}^{(2)}&=&2\int_k \frac{G^2_{k+Q}\Sigma_{k+Q}-G^2_k\Sigma_k}{iQ_0-\ve_{\bk+\bQ}+\ve_\bk}=2\int_k G^2(k)\Sigma(k)\left[\frac{1}{iQ_0-\ve_{\bk}+\ve_{\bk-\bQ}}-\frac{1}{iQ_0-\ve_{\bk+\bQ}+\ve_\bk}\right].
\eea
Now we restrict to the case of small momentum transfers, which is relevant for the Coulomb interaction. In this case, the dispersion of the Green's function in Eq.~(\ref{self}) can be  expanded as $\ve_{\bk+\bq}=\ve_\bk+\bv_\bk\cdot \bq$. The integral of $G^2_k\Sigma_k$ over $k_0$ and $\ve_\bk$ gives, up to a prefactor,
\bea
\int_{\ve_\bk} \int_{k_0}G^2_k\Sigma_k
=-\int_q\int_{\ve_\bk} \int_{k_0} G^2_k G_{k+q}U_q\propto \int_qU_q\frac{\bv_\bk\cdot\bq}{(iq_0-\bv_\bk\cdot\bq)^2}.\label{a5}
\eea
The second factor in the last formula in Eq.~(\ref{a5}) is odd upon $q_0\to-q_0$ and $\bq\to -\bq$, while $U_q$ is even, and therefore $D_{ab}^{(2)}=0$.

Applying Eq.~(\ref{split}) again and using  an explicit form of $\Sigma_k$, we obtain for the remaining part of $D_{ab}$:
\bea
D^{(1)}_{ab}=-2\int_{k,q} \frac{\left(G_k-G_{k+Q}\right)\left(G_{k+q}-G_{k+q+Q}\right)}{(iQ_0-\ve_{\bk+\bQ}+\ve_\bk)^2} U_q.
\eea

Applying  Eq.~(\ref{split})  to diagram {\em c} in Fig.~\ref{fig:diags2}, we obtain
\bea
D_{c}=2\int_{k,q}\frac{\left(G_k-G_{k+Q}\right)\left(G_{k+q}-G_{k+q+Q}\right)}{(iQ_0-\ve_{\bk+\bQ}+\ve_\bk)(iQ_0-\ve_{\bk+\bq+\bQ}+\ve_{\bk+\bq})} U_q.
\eea
Combining the self-energy and exchange contributions, we arrive at the following result for the spin susceptibility
\bea
\chi(Q)&=&D^{(1)}_{ab}+D_{c}=2\int_{k,q} U_{q}\left(G_k-G_{k+Q}\right)\left(G_{k+q}-G_{k+q+Q}\right)R,\label{m11}
\eea
where
\bea
R=\frac{\ve_\bk-\ve_{\bk+\bQ}+\ve_{\bk+\bq+\bQ}-\ve_{\bk+\bq}}{(iQ_0-\ve_{\bk+\bQ}+\ve_\bk)(iQ_0-\ve_{\bk+\bq+\bQ}+\ve_{\bk+\bq})}.
\label{m12}
\eea
By spin conservation, the spin susceptibility must vanish at $\bQ=0$ and finite $Q_0$. Equation (\ref{m11}) satisfies this condition because the numerator of $R$  in Eq.~(\ref{m12})  vanishes at  $\bQ=0$.  After this check, we set $Q_0=0$ upon which $R$ is reduced to
\bea
R=\frac{\ve_\bk-\ve_{\bk+\bQ}+\ve_{\bk+\bq+\bQ}-\ve_{\bk+\bq}}{(\ve_\bk-\ve_{\bk+\bQ})(\ve_{\bk+\bq}-\ve_{\bk+\bq+\bQ})}.\label{R}
\eea
It is clear that $R$ vanishes at finite $\bQ$ and $\bq=0$, which guarantees the gauge invariance of the result.

Expanding the integrand in Eq.~(\ref{m11}) to order $Q^2$, one obtains parts $\delta\tilde\Pi_{s,3}$ and $\delta\tilde\Pi_{s,4}$ of the spin susceptibility, given by Eqs.~(\ref{chi3}) and (\ref{chi4}). Parts $\delta\tilde\Pi_{s,1}$ and $\delta\tilde\Pi_{s,2}$ are absent in this approach as their integrands do not contain factors of $\bq$ and thus vanish identically by gauge invariance.

\section{A $Q^2$ term in the polarization bubble for the static Coulomb potential}
\label{sec:static}
In this Appendix we present details of the calculation of the $Q^2$ term in the polarization bubble for the static Coulomb potential. As discussed in Sec.~\ref{sec:magn},
this is a low-energy contribution, arising from fermions with energies of order of $v_FQ$.

To lowest order in the interaction, the susceptibility is given by the sum of three diagrams in Fig.~\ref{fig:diags2}, where now the wavy line corresponds to a statically screened Coulomb potential; in 2D, $U(q)=2\pi e^2/(q+\kappa)$. In Appendix \ref{app:collect}, we showed how to combine these diagrams in a manifestly gauge-invariant way. The result is presented in Eq.~(\ref{m11}) with $R$ in the static case given by Eq.~(\ref{R}).  Following Ref.~\onlinecite{geldart:1970}, we symmetrize the result by relabeling $\bk+\bq=\bp$, splitting it into two equal parts, and interchanging $\bk\leftrightarrow\bp$ in one of the parts.  Summing also over frequencies $k_0$ and $p_0$, and expanding $\ve_{\bk+\bQ}-\ve_\bk=\bv_\bk\cdot\bQ+{\cal O}(Q^2)$, we obtain
\bea
\delta\Pi(\bQ)= -\int_{\bk,\bp} U(|\bk-\bp|)\left[n_F(\ve_\bk)-n_F(\ve_{\bk+\bQ})\right]\left[n_F(\ve_\bp)-n_F(\ve_{\bp+\bQ})\right] \left[\frac{\bv_\bk.\bQ - \bv_\bp.\bQ}{(\bv_\bk.\bQ) (\bv_\bp.\bQ)} \right]^2.
\eea
The purpose of symmetrization was to make the suppression of the singularity in the Coulomb potential at $\bk=\bp$ more prominent: indeed, now the factor in square brackets vanishes at $\bk=\bp$. However,
the singularity is not removed completely: in the absence of screening ($\kappa=0$), the prefactor of the $Q^2$ term still diverges logarithmically.\cite{rajagopal:1977}

Next, we expand the Fermi functions to third order in $\ve_{\bk+\bQ}-\ve_{\bk}$ and $\ve_{\bp+\bQ}-\ve_{\bp}$, and collect all terms of order $Q^2$. This gives $\delta\Pi(\bQ)=\delta\Pi_1(\bQ)+\delta\Pi_2(\bQ)$, where
\bea
\delta\Pi_1(\bQ)&=& -\frac{1}{4}\int_{\bk,\bp} n_F''(\ve_\bk) n_F''(\ve_\bp)
 \left(\bv_\bk\cdot\bQ - \bv_\bp\cdot\bQ \right)^2U(|\bk-\bp|,\nn\\
\delta\Pi_2(\bQ)&=& -\frac{1}{3}\int_{\bk,\bp} n_F'''(\ve_\bk)n_F'(\ve_\bp) \frac{\bv_{\bk}\cdot \bQ}{\bv_{\bp}\cdot \bQ}
 \left(\bv_\bk.\bQ - \bv_\bp.\bQ \right)^2U(|\bk-\bp|).\label{C2}
\eea
The singularity   at $\bv_\bp\cdot\bQ=0$ in the second line of the above equation is removed by considering the integral in the principal value sense.

At $T=0$, the derivatives of the Fermi functions are reduced to the delta-function and its derivatives, after which it is easy to integrate over $\ve_\bk$ and $\ve_\bp$ by parts. In what follows, we will need
expansions of all factors in the integrand up to ${\cal O}(\ve_\bk^2,\ve_\bp^2)$. These are given by
\bea
k&=&k_F\left[ 1+\frac{\ve_\bk}{2 E_F}-\frac{\ve_\bk^2}{8 E_F^2}+{\cal O}(\ve_\bk^3)\right],\;p=k_F\left[1+\frac{\ve_\bp}{2 E_F}-\frac{\ve_\bp^2}{8 E_F^2}+{\cal O}(\ve_\bp^3)\right];\nn\\
U(|\bk-\bp|)&=& \frac{\pi e^2}{k_F}\left\{ \frac{1}{a+|\sin\frac{\theta}{2}|}- \frac{\ve_\bk + \ve_\bp}{4E_F}\frac{|\sin \frac{\theta}{2}|}{(a+|\sin\frac{\theta}{2}|)^2} + \frac{\ve_\bk^2+\ve_\bp^2}{32 E_F^2}\left[-\frac{\cos\theta}{|\sin\frac{\theta}{2}|(a+|\sin\frac{\theta}{2}|)^2}+ \frac{|\sin\frac{\theta}{2}|}{(a+|\sin\frac{\theta}{2}|)^2}\right.\right. \nn \\
&&+\left.\left.\frac{2\sin^2\frac{\theta}{2}}{(a+|\sin\frac{\theta}{2}|)^3}\right] + \frac{\ve_\bk\ve_\bp}{16 E_F^2}\left[\frac{\cos\theta}{|\sin\frac{\theta}{2}|(a+|\sin\frac{\theta}{2}|)^2}+ \frac{|\sin\frac{\theta}{2}|}{(a+|\sin\frac{\theta}{2}|)^2}
+  \frac{2\sin^2\frac{\theta}{2}}{(a+|\sin\frac{\theta}{2}|)^3}\right]+{\cal O}(\ve_\bk^3,\ve_\bp^3)\right\},\nn\\
  \left(\bv_\bk\cdot\bQ - \bv_\bp\cdot\bQ \right)^2&=& v_F^2Q^2\left[ (\cos\theta_\bk - \cos\theta_\bp)^2 + \frac{1}{E_F}(\cos\theta_\bk-\cos\theta_\bp)(\ve_\bk \cos\theta_\bk - \ve_\bp \cos\theta_\bp)\right.  \nn \\
&&\left.  - \frac{\ve_\bk \ve_\bp}{2E_F^2}\cos\theta_\bk \cos\theta_\bp + \frac{\ve^2_\bk + \ve^2_\bp}{4 E_F^2} \cos\theta_\bk \cos\theta_\bp+{\cal O}(\ve_\bk^3,\ve_\bp^3)\right],
\eea
where $a=\kappa/2k_F$ and $\theta=\theta_\bk-\theta_\bp$.
Integrating by parts over $\ve_\bk$ and $\ve_\bp$ in Eq.~(\ref{C2}), we obtain
\bea
\delta\Pi_1(\bQ) &=& - \frac{ \pi e^2N_F^2 Q^2}{8k_F^3}\int \frac{d\theta_\bk} {2\pi}\int  \frac{d\theta_\bp} {2\pi}\left\{ -\frac{\cos\theta_\bk\cos\theta_\bp}{a+|\sin\frac{\theta}{2}|}+\sin^2\frac{\theta_\bk+\theta_\bp}{2}\left[-\frac{3}{2}\frac{|\sin^3\frac{\theta}{2} |}{(a+|\sin\frac{\theta}{2}|)^2}\right.\right.\nn  \\
&&\left.\left.+ \frac{1}{2}\frac{\cos\theta |\sin\frac{\theta}{2}|}{(a+|\sin\frac{\theta}{2}|)^2}+  \frac{\sin^4\frac{\theta}{2}}{(a+|\sin\frac{\theta}{2}|)^3}\right]\right\},\nn\\
\delta\Pi_2(\bQ)& =& -\frac{\pi e^2 N_F^2 Q^2 }{6k_F^3}\int \frac{d\theta_\bk}{2 \pi} \int \frac{d\theta_\bp}{2\pi} \frac{ \cos\theta_\bk}{ \cos\theta_\bp}
 \left\{ \frac{3\cos^2\theta_\bk-\cos^2\theta_\bp}{2(a+|\sin\frac{\theta}{2}|)}- \frac{\cos\theta|\sin\frac{\theta}{2}|\sin^2\frac{\theta_\bk + \theta_\bp}{2} }{2(a+|\sin\frac{\theta}{2}|)^2}\right.\nn\\
&& \left.+ \frac{\sin^4\frac{\theta}{2}\sin^2\frac{\theta_\bk + \theta_\bp}{2}}{(a+|\sin\frac{\theta}{2}|)^3}
+ \frac{\sin^2\frac{\theta}{2}}{4(a+|\sin\frac{\theta}{2}|)^2}(11\cos\theta_\bk- 3\cos\theta_\bp)\sin\frac{\theta_\bk +\theta_\bp}{2}
\right\}.
\eea
The angular integrals in the equations above are expressed through the angular harmonics
of the following functions
\bea
U_l&=&  \int_0^{2\pi} \frac{d\theta}{2\pi} \cos(l \theta) \frac{1}{a+|\sin\frac{\theta}{2}|}, \nn\\
V_l&=& \int_0^{2\pi} \frac{d\theta}{2\pi} \cos(l \theta)  \frac{\cos\theta |\sin\frac{\theta}{2}|}{(a+|\sin\frac{\theta}{2}|)^2}, \nn\\
W_l&=&\int_0^{2\pi}\frac{d\theta}{2\pi} \cos(l \theta)  \frac{|\sin^3\frac{\theta}{2}|}{(a+|\sin\frac{\theta}{2}|)^2},\nn\\
Z_l&=&  \int_0^{2\pi}\frac{d\theta}{2\pi}   \cos(l \theta)  \frac{\sin^4\frac{\theta}{2}}{(a+|\sin\frac{\theta}{2}|)^3}.
\eea
This gives
\bea
\delta\Pi_1(\bQ)=-\frac{\pi e^2N_F^2Q^2}{16k_F^3}\left(-U_1-\frac 32 W_0+\frac 12 V_0+Z_0\right),\nn\\
\delta\Pi_2(\bQ)=-\frac{\pi e^2N_F^2Q^2}{12k_F^3}\left(\frac 74 U_1-\frac 34U_3+\frac 14 V_0-\frac 12 V_1-\frac 14 V_2-\frac 12 Z_0+Z_1+\frac 12 Z_2 \right).\label{chi12}
\eea
At weak coupling ($a\ll 1$), the harmonics entering the equations above need to be determined to
order ${\cal O}(\ln a)+{\cal O} (1)$. A straightforward computation yields
\bea
U_1&=& \frac{2}{\pi}\left( \ln
\frac{2}{a}-2\right)+{\cal O}\left(a\right),\nn\\
U_3&=& \frac{2}{\pi}\left(\ln
\frac{2}{a}-\frac{46}{15}\right)+ {\cal O}\left(a\right),\nn\\
V_0&=& \frac{2}{\pi}\left(\ln
\frac{2}{a}-3\right)+{\cal O}\left(a\right), \nn\\
V_1&=& \frac{2}{\pi}\left( \ln
\frac{2}{a}-\frac{7}{3}\right)+{\cal O}\left(a\right),\nn\\
V_2&=& \frac{2}{\pi}\left(\ln
\frac{2}{a}-\frac{53}{15}\right)+{\cal O}\left(a\right),\nn\\
W_0 &=&  \frac{2}{\pi} +{\cal O}\left(a\right),\nn\\
Z_0&=&\frac{2}{\pi} +{\cal O}\left(a\right), \nn\\
Z_1&=& -\frac{2}{3\pi}+{\cal O}\left(a^2\right),\nn\\
Z_2&=&-\frac{2}{15\pi} +{\cal O}\left(a^2\right).\label{harm}
\eea
Substituting Eq.~(\ref{harm}) into Eq.~(\ref{chi12}), we obtain Eq.~(\ref{low}) of the main text.
\ewt
\bibliography{q2only.bib}
\end{document}